\documentclass[sn-standardnature,iicol]{sn-jnl}% Standard Nature Portfolio Reference Style
\jyear{2021}%

\theoremstyle{thmstyleone}%
\theoremstyle{thmstyletwo}%

\theoremstyle{thmstylethree}%

\def\apj{\it Astrophys. J.}
\def\apjl{\it Astrophys. J. Lett.}
\def\apjs{\it Astrophys. J. Suppl.}
\def\aj{\it Astron. J.}
\def\mnras{\it Mon. Not. R. Astron. Soc.}
\def\nat{\it Nature}
\def\pasa{\it Publ. Astron. Soc. Austr.}

\def\aap{\it AAP}
\def\prd{\it Phys. Rev. D}
\def\prl{\it Phys. Rev. Lett.}
\def\ssr{\it Space Science Research}
\def\grl{\it Geophysical Research Letters}

\newcommand{\gtorder}{\mathrel{\raise.3ex\hbox{$>$}\mkern-14mu
            \lower0.6ex\hbox{$\sim$}}}
\newcommand{\ltorder}{\mathrel{\raise.3ex\hbox{$<$}\mkern-14mu
            \lower0.6ex\hbox{$\sim$}}}

\begin{document}

\title[kHz QPOs in sGRBs]{Kilohertz quasiperiodic oscillations in short gamma-ray bursts}

\author*[1,2,3,4]{\fnm{Cecilia} \sur{Chirenti}}%\email{chirenti@umd.edu}

\author[5]{\fnm{Simone} \sur{Dichiara}}%\email{sbd5667@psu.edu}

\author[6]{\fnm{Amy} \sur{Lien}}%\email{alien@ut.edu}

\author[1]{\fnm{M. Coleman} \sur{Miller}}%\email{miller@astro.umd.edu}

\author[7]{\fnm{Robert} \sur{Preece}}%\email{preecer@uah.edu}

\affil*[1]{\orgdiv{Department of Astronomy}, \orgname{University of Maryland}, \orgaddress{\city{College Park}, \postcode{20742}, \state{MD}, \country{USA}}}

\affil*[2]{\orgdiv{Astroparticle Physics Laboratory}, \orgname{NASA/GSFC}, \orgaddress{\city{Greenbelt}, \postcode{20771}, \state{MD}, \country{USA}}}

\affil*[3]{\orgdiv{Center for Research and Exploration in Space Science and Technology}, \orgname{NASA/GSFC}, \orgaddress{\city{Greenbelt}, \postcode{20771}, \state{MD}, \country{USA}}}

\affil*[4]{\orgdiv{Center for Mathematics, Computation and Cognition}, \orgname{UFABC}, \orgaddress{\city{Santo Andre}, \postcode{09210-170}, \state{SP}, \country{Brazil}}}

\affil[5]{\orgdiv{Department of Astronomy and Astrophysics}, \orgname{The Pennsylvania State University}, \orgaddress{\street{525 Davey Lab}, \city{University Park}, \postcode{16802}, \state{PA}, \country{USA}}}

\affil[6]{\orgdiv{Department of Chemistry, Biochemistry, and Physics}, \orgname{University of Tampa}, \orgaddress{\street{401 W. Kennedy Blvd}, \city{Tampa}, \postcode{33606}, \state{FL}, \country{USA}}}

\affil[7]{\orgdiv{Department of Space Science}, \orgname{University of Alabama in Huntsville}, \orgaddress{\city{Huntsville}, \postcode{35899}, \state{AL}, \country{USA}}}

%%==================================%%
%% unstructured abstract                                     %%
%%==================================%%

\abstract{Short $\gamma$-ray bursts are associated with binary neutron star mergers, which are multimessenger astronomical events that have been observed both in gravitational waves and in the multiband electromagnetic spectrum \cite{2017ApJ...848L..12A}. Depending on the masses of the stars in the binary and on details of their largely unknown equation of state, a dynamically evolving and short-lived  neutron star may be formed after the merger, existing for approximately 10-300 ms before collapsing to a black hole \cite{2000ApJ...544..397S,2012PhRvD..86f4032P}. Numerical relativity simulations across different groups consistently show broad power spectral features in the 1-5 kHz range in the post-merger gravitational wave signal \cite{2005PhRvD..71h4021S,2005PhRvL..94t1101S,2008PhRvD..78b4012L,2008PhRvD..78h4033B,2013PhRvD..88d4026H,2014PhRvL.113i1104T,2015PhRvD..91f4001T,2015PhRvD..91l4041D,2016ApJ...824L...6R,2017ApJ...842L..10R,2022PhRvL.128p1102B}, which is inaccessible by current gravitational-wave detectors but could be seen by future third generation ground-based detectors in the next decade \cite{2010CQGra..27s4002P,2017CQGra..34d4001A,2020PASA...37...47A}. This implies the possibility of quasiperiodic modulation of the emitted $\gamma$-rays in a subset of events where a neutron star is formed shortly prior to the final collapse to a black hole \cite{2019ApJ...884L..16C,2019LRR....23....1M,2020ApJ...901L..37M,2021ApJ...906..127F}. Here we present two such signals identified in the short bursts  GRB~910711 and GRB~931101B from archival BATSE data, which are compatible with the predictions from numerical relativity.}

%\keywords{Neutron star, gamma-ray burst, gravitational wave, numerical relativity, equation of state}



\maketitle

\begin{figure}[h!]
\centering
\includegraphics[width=0.45\textwidth]{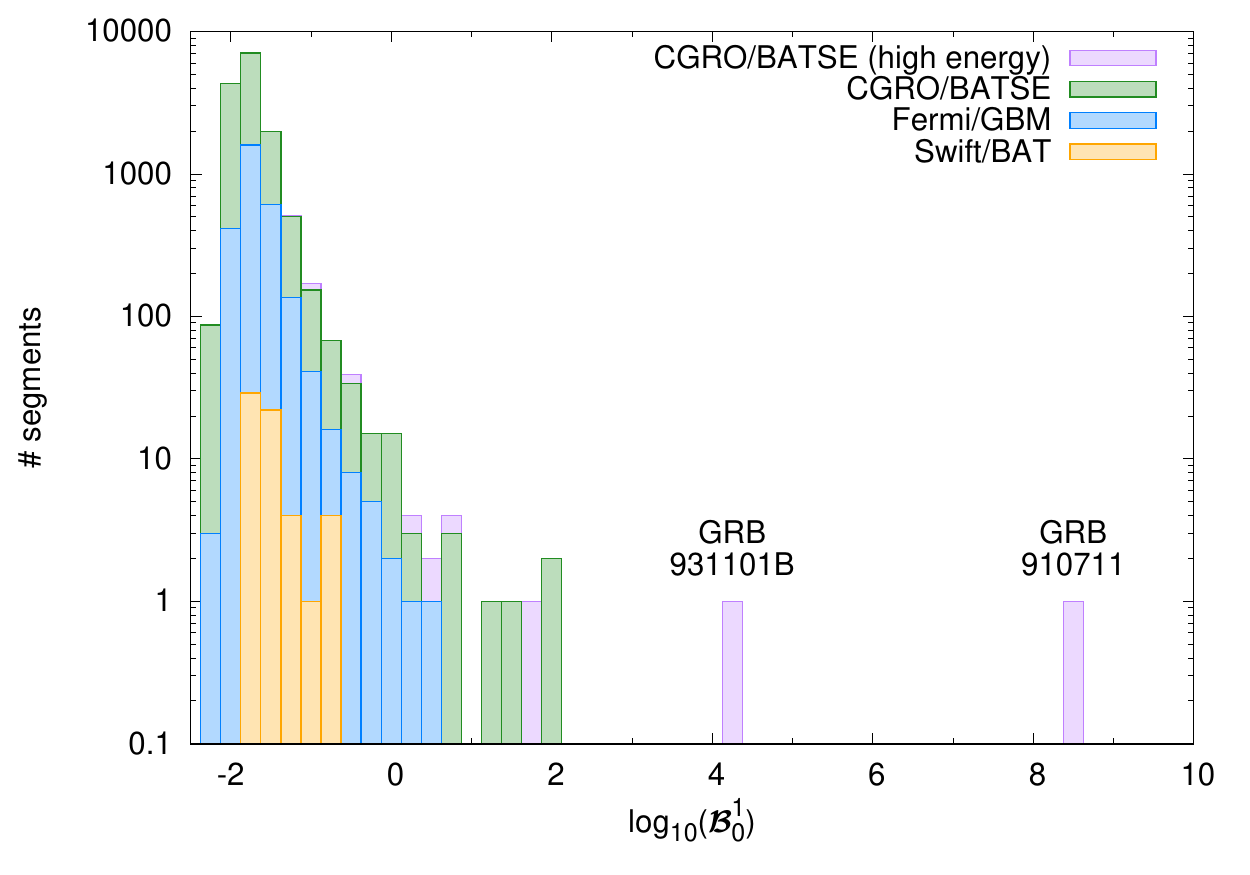}
\caption{$\vert$ \textbf{Differential distribution of Bayes factors.}  Here we plot, for different sets of $\sim 0.1$-second segments of short bursts, the $\log_{10}$ of the Bayes factor ${\cal B}_0^1$ between a model with one Lorentzian QPO plus white noise, and a model with just white noise, in the frequency range $500-5000$~Hz \cite{2019ApJ...871...95M}.  In orange we show the distribution for Swift/BAT bursts, in blue for Fermi/GBM bursts, in green  for CGRO/BATSE bursts when we sum the counts over all four TTE energy channels, and in purple for CGRO/BATSE bursts when we sum only the counts from the two highest TTE energy channels (i.e., energies $>100$~keV).  In each case we have cleaned the sample by removing segments contaminated by cosmic ray spikes, excess red noise, or other features which artificially increase the rate of false positives.  Most of these segments are consistent with noise, but the two outliers on the BATSE high energy distribution (in purple, extending to the right) have overwhelmingly larger ${\cal B}_0^1$ than the rest.  These are the signals that we investigate.}
\label{fig:Bayes}
\end{figure}

\begin{figure}[h!]
\centering
\includegraphics[width=0.45\textwidth]{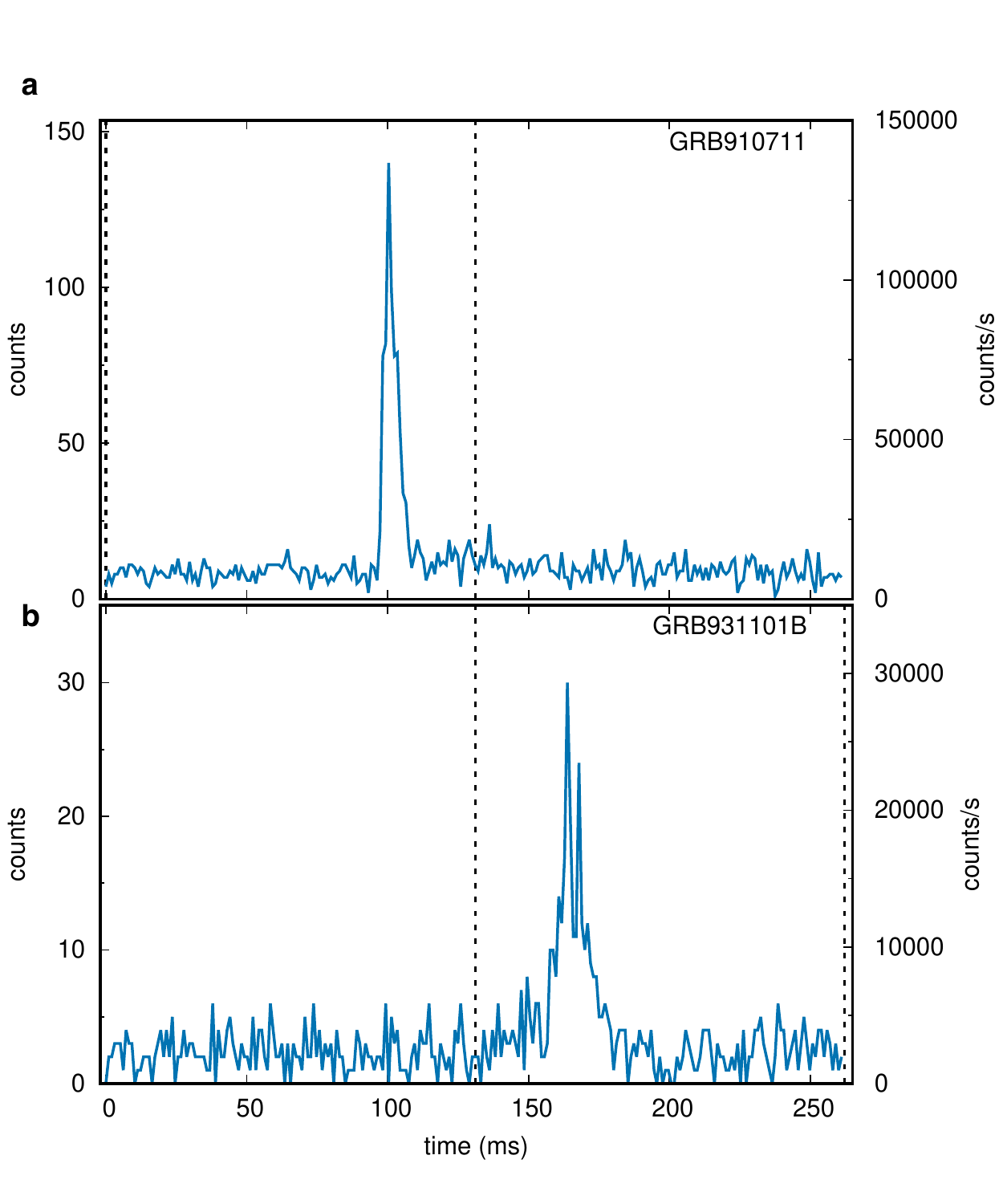}
\caption{$\vert$ \textbf{Light curves of the two bursts with signals.} \textbf{a,} Counts per 1.024-millisecond bin in the two highest-energy channels (channels 3 and 4) in the BATSE TTE data, for consecutive 1.024-millisecond intervals beginning at the start of the TTE data for GRB~910711. \textbf{b,} The same, for GRB~931101B. The signals were found in the segments bracketed by the vertical dotted lines.}
\label{fig:lightcurves}
\end{figure}

\begin{figure}[h!]
\centering
\includegraphics[width=0.45\textwidth]{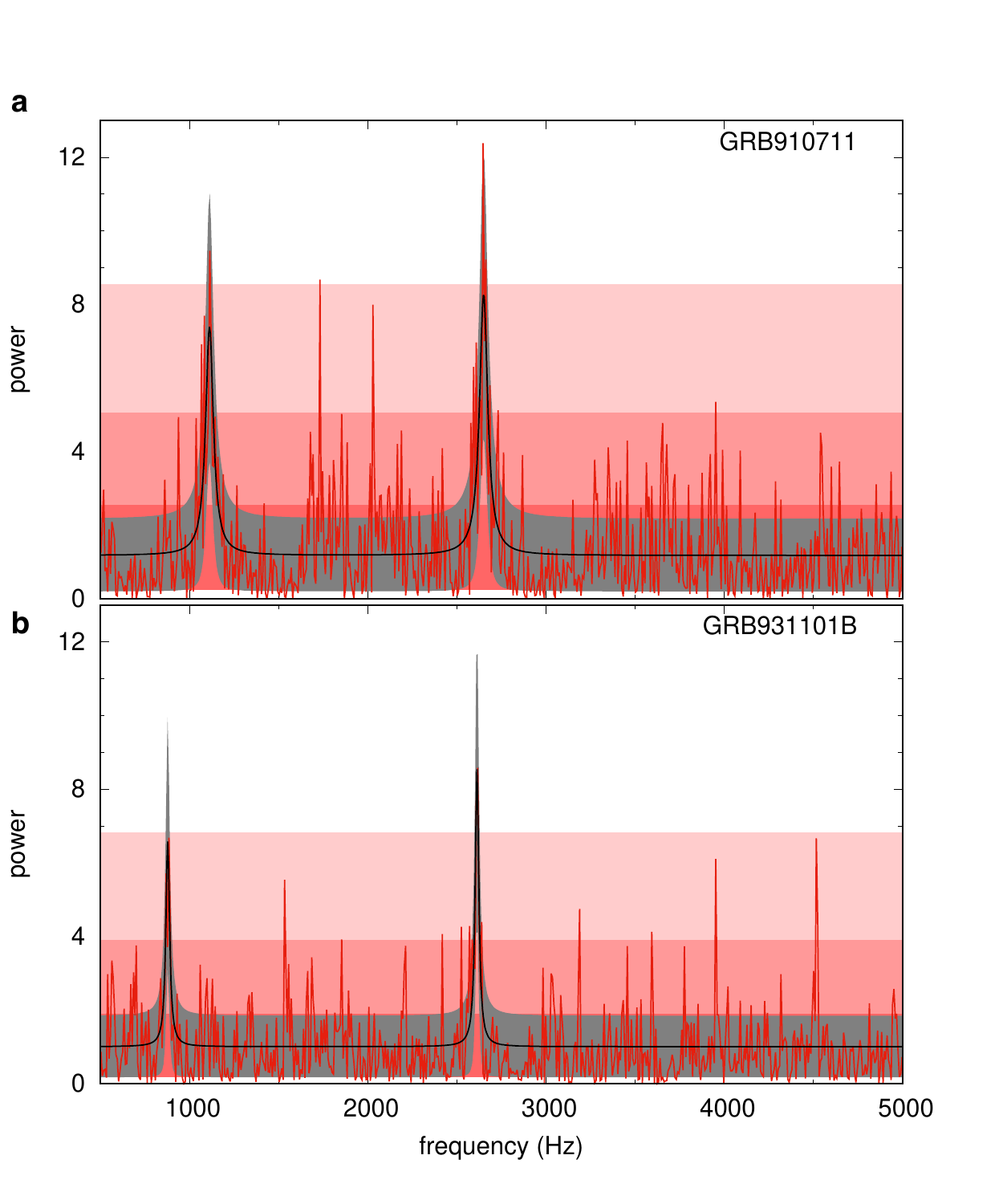}
\caption{$\vert$ \textbf{Power spectra of the two bursts with signals.} \textbf{a,} Power spectrum of GRB~910711.  \textbf{b,} Power spectrum of GRB~931101B. Here we use the intervals delineated by the vertical dashed lines in Figure~\ref{fig:lightcurves}. These each have a duration of 0.131072 seconds and thus the frequency resolution of the power spectrum is $1/0.131072~{\rm s}=7.6294$~Hz.  We use the Groth power normalization \cite{1975ApJS...29..285G}, in which the expected power averages 1 if the flux is intrinsically constant and has only photon counting (Poisson) noise.  In addition to the power densities (red lines) we also show the $1\sigma, 2\sigma$ and $3\sigma$ single-trial power ranges for the best white noise only fits in each case (red bands) and the $\pm 1\sigma$ range for the power expected in the best two-QPO plus white noise fits (grey bands). The corresponding best-fit values are shown in Table \ref{tab:bestfits} in Methods. The range of frequencies shown in the figure, $500-5000$~Hz, is what we use in our QPO search and is intended to reach the highest plausible oscillation frequencies but to avoid red noise at low frequencies.  We see that the two bursts have similar power density structures, with clear peaks at $\sim 2600$~Hz and $\sim 1000$~Hz. }
\label{fig:powerspect}
\end{figure}

\begin{figure}[h!]
\centering
\includegraphics[width=0.45\textwidth]{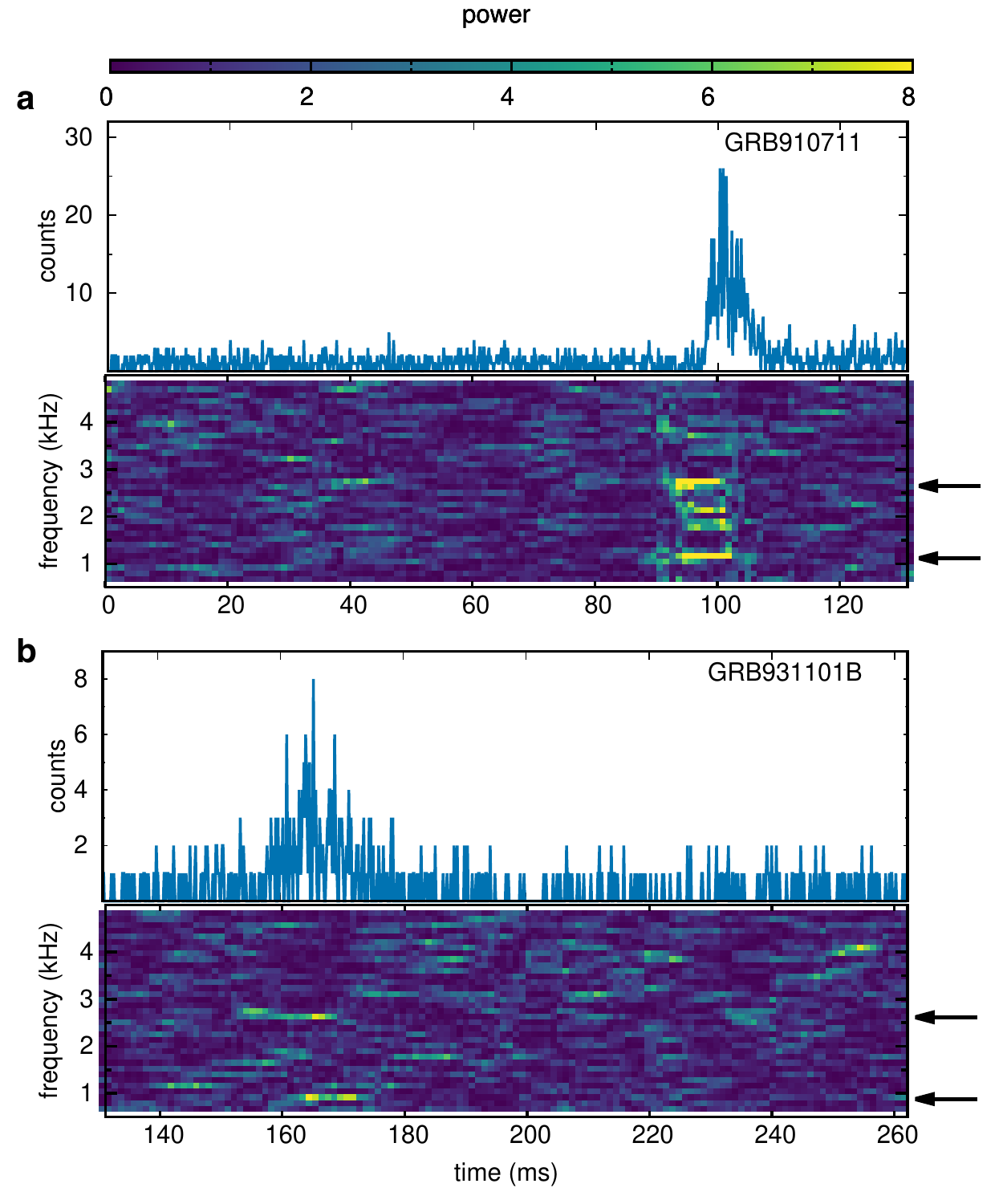}
\caption{$\vert$ \textbf{Spectrograms for the burst segments with signals.} \textbf{a,} Spectrogram for GRB~910711.  \textbf{b,} Spectrogram for GRB~931101B. Here we use the parts of the bursts bracketed by the vertical dotted lines in Figure~\ref{fig:lightcurves}.  For each burst, the top panel shows the light curve binned in 128-microsecond intervals (in contrast with the 1.024-millisecond intervals of Figure~\ref{fig:lightcurves}).  Each pixel shows the power (indicated by the color bar) at the associated frequency for an 8.192-msec interval, with a start time in milliseconds given on the horizontal axis. The black arrows indicate the mean values of the QPO frequencies given in Table \ref{tab:burstqpos}. The maximum power in a pixel is $\sim 16$ for GRB~910711 and $\sim 8$ for GRB~931101B.  Note that in GRB~910711 there is substantial power at the main $\sim 2600$~Hz frequency some $0.08-0.06$~seconds before the burst, which could indicate precursor emission.}
\label{fig:spectrogram}
\end{figure}

Given the anticipated high frequencies, we analyzed data from gamma-ray observatories with excellent time resolution: the Fermi Gamma-ray Burst Monitor (GBM) \cite{2009ApJ...702..791M}; the Burst Alert Telescope (BAT) on the Neil Gehrels Swift Observatory \cite{2005SSRv..120..143B}, and the Compton Gamma-ray Observatory (CGRO) Burst and Transient Source Experiment (BATSE) in Time Tagged Event (TTE) mode \cite{2000ApJS..126...19P}.  Previous searches for periodic signals in gamma-ray bursts yielded null results \cite{1997ApJ...491..720D,2002ApJ...576..932K,2013ApJ...777..132D}, but the dynamical nature of the merger suggests that instead of periodic signals, quasiperiodic oscillations (QPOs) are more probable.  We focused on short gamma-ray bursts based on the expectation that these are due to neutron star mergers and thus could have an oscillatory phase, e.g., as a hypermassive neutron star (HMNS), which temporarily avoids collapse to a black hole due to the star's differential rotation \cite{2021GReGr..53...59S}.  Figure~\ref{fig:Bayes} shows that in these data sets, two BATSE bursts stand out with strong signals. Figures~\ref{fig:lightcurves}, \ref{fig:powerspect}, and \ref{fig:spectrogram} show respectively the light curves, power spectra, and spectrograms for these two bursts, and Table~\ref{tab:burstqpos} gives the properties of their QPOs.  We find (see Methods for details) that when all trials factors are taken into account, the probability that the combined catalogs of BATSE, GBM, and BAT would produce false positive QPOs of the strength we observe is $\sim 3\times 10^{-7}$. 

%\newpage

Both of our signals (out of more than 700 total bursts; see Methods) are in BATSE bursts, which is to be expected because BATSE has a larger detector area than GBM or BAT, which makes it easier to detect modulations in the count rate.  This may suggest that future large-area instruments with excellent time resolution, such as STROBE-X \cite{2019arXiv190303035R} or AMEGO-X \cite{2022HEAD...1940403C}, will identify burst QPOs that are currently too weak to detect.

The frequencies of the QPOs in our two featured bursts are broadly consistent with the expectations from numerical relativity simulations of double neutron star mergers.  Possible detections of kilohertz QPOs have also been reported in giant flares from two soft gamma-ray repeaters (SGRs) \cite{2006ApJ...653..593S,2021Natur.589..207R}, but
the high implied isotropic energies and luminosities of GRBs 910711 and 931101B argue against the SGR giant flare interpretation for our bursts (see Methods).  We therefore adopt the hypothesis that these are classical short gamma-ray bursts resulting from the merger of two neutron stars.  Even if this is the case it does not necessarily follow that the QPOs we observe come from the oscillations of an HMNS.  It is, for example, conceivable that the oscillations come from a lower-mass neutron star or from some properties of accretion onto a black hole in the so-called lower mass gap ($2 - 5 M_{\odot}$); further modeling will as always be necessary.  

If the high-frequency QPOs that we detect are indeed related to HMNS oscillations, then the frequencies detected in our signals can be compared with several phenomenological relations identified for the frequencies observed in the simulated post-merger gravitational waveforms from binary neutron star mergers \cite{2015PhRvD..91f4001T,2015PhRvD..91l4056B,2015PhRvD..92l1502P,2015PhRvD..91f4027K,2017PhRvD..95f3016C}.  For instance, if the unknown redshift $z$ of the sources can be neglected, the frequencies $\nu_2$ of the main peak presented in Table \ref{tab:burstqpos} together with a phenomenological relation obtained from simulations of mergers of two 1.35 $M_{\odot}$ neutron stars \cite{2021PhRvD.104d3011L} suggest a radius $R_{1.6} \simeq 13$ km for a $1.6~M_{\odot}$ neutron star. It is unlikely that such a bright burst as GRB 910711 happened at a large redshift, but the redshift correction would mean that the rest-frame frequency would be higher by a factor of $1+z$, and thus the inferred radius would be smaller. The best estimate of the radius decreases linearly with small $z$; e.g., $R_{1.6} \simeq 12.5$ km for $z = 0.1$ and $R_{1.6}\simeq 12.0$ km for $z=0.2$. 

Another possible inference is that of the spin of the HMNS. It is currently understood that HMNSs are supported against gravitational collapse by rapid and differential rotation. During their brief lifetime, they should be the fastest rotating stars known. Simulations show that the frequency $\nu_2$ corresponds to twice the maximum angular velocity inside the star \cite{2015PhRvD..91f4027K}, $\Omega_{\rm max}$. Our results would imply $\Omega_{\rm max} \sim (1+z)1.3\, {\rm kHz}$ (allowing for a redshift correction), which even at $z = 0$ is almost two times higher than any neutron star spin frequency observed to date, and above the expected Keplerian mass-shedding limit for a uniformly rotating HMNS, therefore being consistent with the need for differential rotation.

Additional information could presumably be provided by identification of the lower-frequency peak $\nu_1$, but this is complicated by uncertainties in the identification of the peak in different models as well as the difficulty of accurate inclusion of fully resolved neutrino physics, magnetohydrodynamic turbulence, and so on in long-term numerical simulations.  Moreover, because we observe gamma rays rather than gravitational waves, additional careful modeling of the jet is needed to draw a connection between gravitational waveforms and the observed modulations of gamma rays. For example, fluctuations in the post-merger accretion disk can also drive changes in the jet and thus the observed GRB flux \cite{2006MNRAS.368.1609D,2019ApJ...886L..30N}.  Nonetheless, the detection of these high-frequency QPOs provides a potentially powerful new tool to study the dynamics and gravity of merging neutron stars.

%\newpage 
\begin{table*}[h!]
\begin{center}
\begin{minipage}{\textwidth}
\caption{$\vert$ \textbf{Bursts with QPOs, Bayes factors and median and $\pm 1\sigma$ values of central frequencies and widths}}\label{tab:burstqpos}
\setlength{\tabcolsep}{2pt}
\begin{tabular}{lcccccccc}
\toprule
GRB & $T_{90}$ (msec) & Counts & ${\cal B}_0^1$ & ${\cal B}_0^2$ & $\nu_1 ({\rm Hz})$ & $\Delta\nu_1$ (Hz) & $\nu_2$ (Hz) & $\Delta\nu_2$ (Hz) \\
\midrule
910711  & 14 \cite{1999ApJ...527..827C} & 1790 & $6.5\times 10^8$ & $3.0\times 10^{17}$ & $1113^{+7}_{-8}$ & $25^{+9}_{-7}$ & $2649^{+6}_{-7}$ & $26^{+9}_{-7}$\\
931101B & 34 \cite{1999ApJ...527..827C} & 524 & $1.9\times 10^4$ & $3.6\times 10^6$ & $877^{+6}_{-8}$ & $15^{+7}_{-2}$ & $2612^{+9}_{-8}$ & $14^{+7}_{-3}$\\
\bottomrule
\end{tabular}
\footnotetext{Note: The frequencies $\nu$ and half widths at half maximum $\Delta\nu$ for the QPOs are from the two-QPO plus white noise fit to the data from 500~Hz through 5000~Hz for a 0.131072-second segment of BATSE TTE data. (The one-QPO plus white noise fit to the data, corresponding to Figure \ref{fig:Bayes}, finds only the second QPO listed here.)  We analyze data only from the highest two of the four BATSE TTE energy channels, and list the total number of counts in those two channels over the 0.131072-second segments that contain the signals.  $T_{90}$ is the shortest time span that contains 90\% of the burst counts and ${\cal B}_0^1$ (${\cal B}_0^2$) is the Bayes factor between a one-QPO (two-QPO) plus white noise model and a white noise only model. For the frequencies and frequency widths we give the median and the $\pm 1\sigma$ ranges.}
\end{minipage}
\end{center}
\end{table*}


%\newpage 
\section*{Methods}

\renewcommand{\tablename}{Extended Data Table}
\renewcommand{\thetable}{ \arabic{table}}
\setcounter{table}{0}

\bmhead{Other Bayesian searches for high-frequency QPOs in GRBs}
\label{sec:others}
Other searches have been conducted, with null results, for high-frequency variability in gamma-ray burst data \cite{1997ApJ...491..720D,2002ApJ...576..932K,2013ApJ...777..132D}.  For example, an analysis of BATSE TTE data from 20 bright short and long bursts (not including either of the two we feature here) found no significant periodic signals in the frequency range 30~Hz to 60,000~Hz \cite{1997ApJ...491..720D}. Later analyses of BATSE TTE data for periodic signals in the $400-2500$~Hz range using a Rayleigh test \cite{2002ApJ...576..932K}, and for oscillations and narrow quasiperiodic oscillations of up to $200$~Hz using a Fourier-based method \cite{2013ApJ...777..132D}, also found no significant signals (the approach \cite{2010MNRAS.402..307V} in the latter study was also used to look for lower-frequency QPOs in short magnetar bursts and long GRBs \cite{2013ApJ...768...87H,2014ApJ...787..128H,2014ApJ...795..114H,2016A&A...589A..98G}).  In contrast, we search for the broad quasi-periodicity expected from a temporary, decaying oscillation.

\bmhead{Data selection}
\label{sec:selection}
In order to select bursts for our analysis that could be bright enough to yield detections or interesting limits on a QPO signal, we use a flux threshold for the Swift/BAT and Fermi/GBM bursts based on theoretical arguments \cite{2019ApJ...884L..16C}. Using Fourier analysis \cite{1988SSRv...46..273L}, we can estimate a flux threshold
\begin{equation}
    F \approx \frac{2 n_{\sigma}E_{\rm peak}}{A_{\rm det}a_{\rm osc}^2}\sqrt{\frac{\Delta \nu}{\Delta T}}\,,
\label{eq:fluxthresh}
\end{equation}
where $n_{\sigma}$ is the required significance of the QPO (in standard deviations), $E_{\rm peak}$ is the peak energy of the burst, $A_{\rm det}$ is the detector area, $a_{\rm osc}$ is the fractional root mean squared oscillation amplitude in the count rate caused by the QPO, $\Delta T$ is the total observation time and $\Delta \nu$ is the width of the QPO. We require $n_{\sigma} = 5$ and fix a fiducial high $a_{\rm osc} = 0.75$ to obtain a correspondingly low $F$, and assumed $\Delta\nu=250$~Hz \cite{2019ApJ...884L..16C}.  In contrast to this selection for the Swift/BAT and Fermi/GBM short bursts, we analyzed all of the $532$ short bursts from the BATSE sample. (Due to its larger detector area, the BATSE flux threshold would be up to several times lower than the threshold for the other detectors.)

\bmhead{Swift/BAT data set}
\label{sec:swift}
In this analysis, we use a sub-sample of 8 short GRBs ($T_{90} < 5\,{\rm s}$) and 110 long GRBs (included here for improved statistics)  detected with Swift/BAT (before August 10, 2020) with a 1-s peak flux in the 15-150 keV band higher than $3.556 \times 10^{-7} \ \rm erg \ s^{-1} \ cm^{-2}$ based on Equation~\ref{eq:fluxthresh}. The 1-s peak fluxes are adopted from the Swift BAT/GRB catalog \cite{2016ApJ...829....7L}. The sample excludes two ground-detected GRBs that were found during spacecraft slews, which could complicate the analysis due to the continuously changing instrumental response and sky coverage during the slew time throughout the entire burst emission.

Our QPO study utilizes BAT non-maskweighted light curves in the 15-350 keV range and uses the native 100 $\mu$s time resolution for our intervals. We applied the standard HEASoft tool, batbinevt (version 1.48), to create the light curves. For each burst, we search for QPO signals across the entire burst duration.

\bmhead{Fermi/GBM data set}
\label{sec:fermi}
We selected a sub-sample of 184 short GRBs ($T_{90} < 5\,{\rm s}$) detected by Fermi-GBM from July 2008 to July 20, 2018. We require a peak flux higher than 2.074$\times 10^{-6}$~erg~cm$^{-2}$~s$^{-1}$ based on Equation~\ref{eq:fluxthresh}. 
For each GRB we take the TTE files of the two most illuminated NaI detectors and extract the corresponding light curves in 100 $\mu$s intervals (binned from the native $2\, \mu$s resolution) in the 8–1000 keV band using the \textsc{fselect} and \textsc{gtbin} tools. For our analysis we use the \textsc{heasoft} (version 6.30.1) and the \textsc{Fermitools} (version 2.0.8) software packages. 
We reject light curves affected by spikes due to the interactions of high-energy particles with the spacecraft \citep{2009ApJ...702..791M} and use non-background subtracted data. 
We process the data by following the Fermi team threads \citep{FermiAnalysis}.

\bmhead{CGRO/BATSE data set}
\label{sec:batse}
We analyze the BATSE TTE data available for all 532 short gamma-ray bursts \citep{BATSETTE}, without a lower flux threshold. In this mode, the time resolution is 2 $\mu$s and there are four energy channels: channel 1 has photon energies from $20-50$~keV, channel 2 from $50-100$~keV, channel 3 from $100-300$~keV, and channel 4 has energies $>300$~keV \citep{BATSE3b}. The TTE data contain 32,768 counts covering no more than 2 s, with about one fourth of the counts in the preburst interval, which contains data from all 8 detectors. The remaining 3/4 of memory contains data for burst-selected detectors (brightest detectors at the trigger time).

\bmhead{Analysis Methods}
\label{sec:analysis}
We search short GRBs for QPOs that are reasonable matches to the expectations for post-merger oscillations of HMNSs.  Because these oscillations are expected to have frequencies in excess of $\sim 1000$~Hz, we focus on high frequencies: our analysis uses power density spectra from 500~Hz to 5000~Hz.  In addition, because HMNS oscillations are expected to damp on timescales $\sim 0.1$~s \cite{2000ApJ...544..397S,2012PhRvD..86f4032P}, we construct power density spectra from data segments of approximately the same duration.  

In practice, since we use a fast Fourier transform (FFT), the segments we analyze are powers of 2 times the time resolution of each instrument. We therefore use segments of length $2^{10}\times 10^{-4}$~s, or $0.1024$~s, for the Swift BAT and Fermi GBM data sets, which have Nyquist frequencies of $1/(2\times 100\,\mu {\rm s})=5000$~Hz.  For BATSE data sets we analyze segments of length $2^{16}\times 2\times 10^{-6}$~s, or $0.131072$ s.  The Nyquist frequency for these data is $1/(2\times 2\, \mu {\rm s})$, or $250,000$~Hz, but at least in our two featured bursts we did not see any excess power beyond 5000~Hz.

In order to avoid missing signals that overlap the end of one segment and the beginning of the next, our segments have half-overlap.  For example, if a Swift burst lasts for two seconds, our first segment is from 0 to 0.1024 s, our second is from 0.0512 to 0.1536 s, our third is from 0.1024 to 0.2048 s, and so on.  We have not attempted to optimize the segments further; for example, if there was an apparent signal from 0.0512 to 0.1536 s in a given data set, we did not explore slightly offset intervals of the same duration.

When we search for QPOs we compute Bayes factors between specified models.  In the high frequency range that we explore, we typically do not expect (nor do we typically find) that there is significant red noise (see the ``Additional analysis with adjustable red noise" section below for red-noise analysis of our two signals, which finds no significant difference in the significance or inferred parameters when we use red instead of white noise).  However, very short-duration pulses such as those that could be produced by a cosmic ray will generate noise that is close to white because the Fourier transform of a delta function is a constant with frequency.  As a result, our no-QPO model is that, in addition to the unavoidable Poisson fluctuations, there can be additional white noise.  

For our QPO model we use a Lorentzian form for the power density $P(\nu)$, which we add to white noise.  Therefore, in a model with $n$ QPOs the power as a function of frequency $\nu$ is
\begin{equation}
P(\nu)=A_{\rm white}+\sum_{i=1}^n\frac{A_i}{1+(\nu-\nu_i)^2/(\Delta \nu_i)^2}\; .
\end{equation}
Thus a model with only white noise has 1 parameter; a model with white noise plus one QPO has four parameters; a model with white noise plus two QPOs has seven parameters; and so on.
For QPO $i$, $A_i$ is the power density of the Lorentzian at its center, $\nu_i$ is the central frequency of the Lorentzian, and $\Delta \nu_i$ is the half-width of the Lorentzian.  We adopt this form because it is the Fourier transform of a signal described by 
an exponentially damped sinusoid, which is roughly consistent with the expectations for a decaying HMNS oscillation.  Each QPO is represented by a different Lorentzian, and if the frequencies of multiple QPOs are close enough relative to their widths, they can overlap.  In our QPO models we also allow there to be additional white noise.

Details about our approach have been published earlier \cite{2019ApJ...871...95M}.  In brief, the likelihood ${\cal L}(P; P_s)$ that a power density $P$ will be observed in a given frequency bin, given an expected model power $P_s$, is \cite{1975ApJS...29..285G}:
\begin{equation}
{\cal L}(P; P_s)=e^{-(P+P_s)}\sum_{m=0}^\infty \frac{P^m P_s^m}{(m!)^2}\; .
\end{equation}
Here the normalization is such that if the signal is intrinsically constant and thus the only power comes from Poisson fluctuations (i.e., $P_s=0$), the mean power is $\langle P\rangle=1$.  Note that this differs by a factor of two from the commonly-used Leahy normalization \citep{1983ApJ...266..160L}, for which $\langle P\rangle=2$ for pure Poisson noise.  For a given power density spectrum and a given model (involving 0, 1, or 2 QPOs), we compute the log of the likelihood of the data set given the model by summing the log likelihoods over all of the power densities from 500~Hz to 5000~Hz.  

For each model type (0, 1, or 2 QPOs) we compute the maximum log likelihood over all parameter combinations using a custom affine-invariant Markov chain Monte Carlo (MCMC) code based on the approach of Goodman and Weare \cite{2010CAMCS...5...65G}.  In our particular implementation of the MCMC code we use 32 walkers for the white noise only and the white noise + 1 QPO runs, and 56 walkers for the white noise + 2 QPOs runs.  We start each run from a tight bundle of walkers near a random point in parameter space, and perform each run 20 times from different random starting locations.  We find that similar values for the best fit are repeatedly attained in these 20 runs for a given data set, which suggests that the likelihood surfaces are relatively smooth.

To compute Bayes factors we need first to compute the evidence $E$ for each model by integrating the product of the likelihood ${\cal L}({\vec a})$ with the prior $q({\vec a})$ over all combinations of the parameters ${\vec a}$ in a given model:
\begin{equation}
E=\int {\cal L}({\vec a})q({\vec a})d{\vec a}\; ,
\end{equation}
where the prior has been normalized such that $\int q({\vec a})d{\vec a}=1$.  We calculate $E$ using Monte Carlo integration, with a target precision of 10\% of the best estimate.  The Bayes factor between models $A$ and $B$ is then ${\cal B}^A_B=E^A/E^B$.  We assume that the 0-, 1-, and 2-QPO models all have the same prior probability, which means that the odds ratio between them equals the Bayes factor: ${\cal O}^A_B={\cal B}^A_B$.

Based on our experience with the data, we use the following priors:

\begin{enumerate}
\item For the white noise, $A_{\rm white}$ is flat between 0 and 5.

\item For the QPOs, $A_i$ is flat between 0 and 30.

\item For the higher-frequency QPO, $\log_{10}\nu_2({\rm Hz})$ is flat between 3.0 and 3.7.  For the lower-frequency QPO, $\log_{10}\nu_1({\rm Hz})$ is flat between 2.7 and 3.7. For a 1-QPO model we use only the $\nu_2$ prior.

\item $\log_{10} \Delta \nu_i({\rm Hz})$ is flat between 1.0 and 3.0.
\end{enumerate}

We adopt priors for the centroid frequency and width of the QPOs that are flat in log frequency because we wish to be agnostic about the scale; in contrast, for example, a prior that is flat in frequency width between 10~Hz and 1000~Hz would have most of its weight at large widths.  However, the strength of the QPO signal in our two featured bursts is great enough that the prior has little effect.

\bmhead{Potential causes of false QPO signals}
\label{sec:false}
Our initial analysis, which began with the Swift/BAT data, revealed some possible causes of false QPO signals.  The first is spikes of $\sim 10-100$ counts within a single readout time interval of $100~\mu{\rm s}$.  These are presumably produced by cosmic rays rather than gamma rays from a burst.  As discussed above, because an unresolved spike approximates a delta function, the power density from a single spike is essentially white noise.  If there are two or more such spikes in a given data set then the frequencies corresponding to the reciprocals of the intervals between the spikes also show up prominently in the power density spectrum, and these can be read incorrectly as QPOs.   Our approach was to discard segments of Swift/BAT data in which any $100~\mu{\rm s}$ interval had more than 8 counts, because such a high number is almost certainly from a cosmic ray (see the orange distribution in Extended Data Figure~\ref{fig:B_Swift}). 
\renewcommand{\figurename}{Extended Data Fig.}
\renewcommand{\thefigure}{ \arabic{figure}}

\setcounter{figure}{0}

\begin{figure}[h!]
\centering
\includegraphics[width=0.45\textwidth]{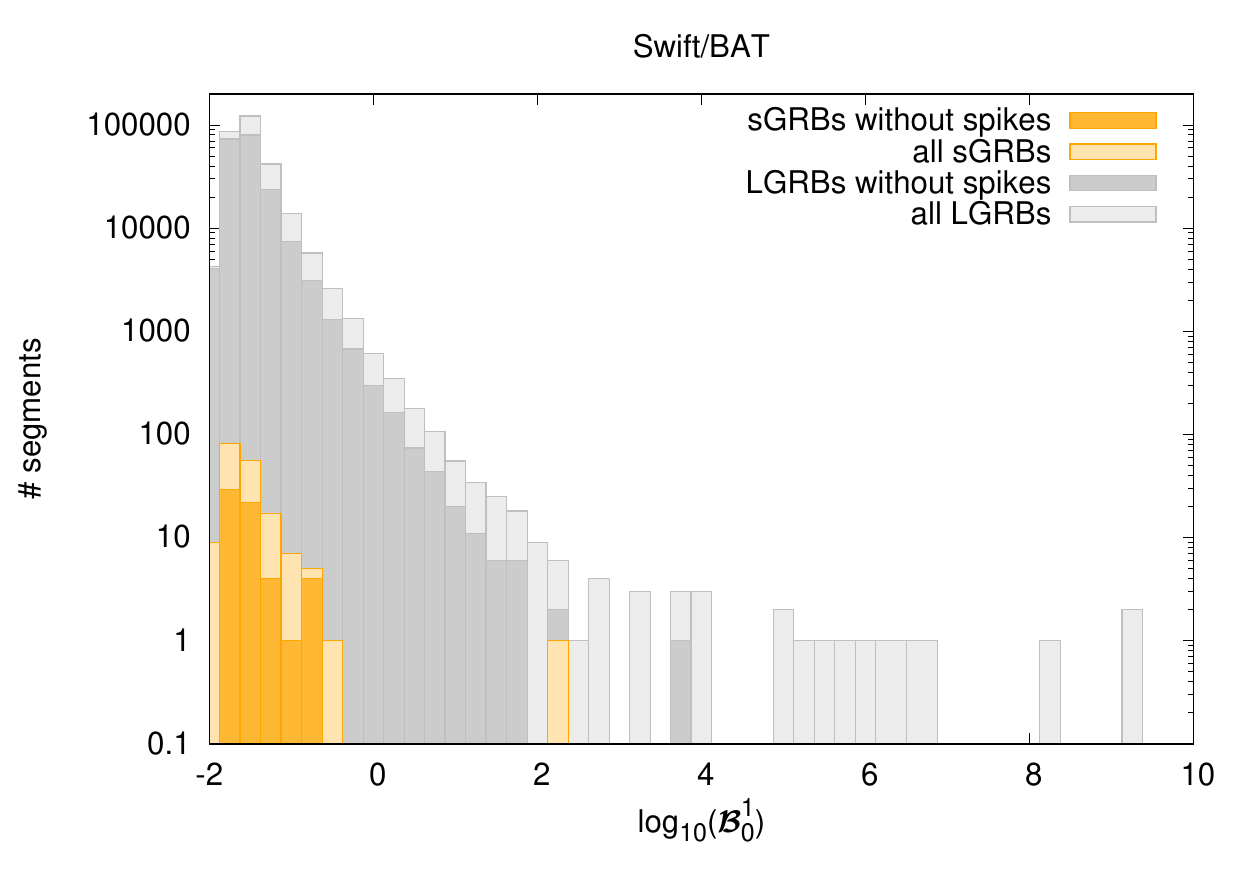}
\caption{$\vert$ \textbf{Differential distribution of Bayes factors for Swift/BAT bursts.} We plot separately the sample of 8 short bursts (orange) and 110 long bursts (gray). The light orange and light gray distributions include the analysis of segments with spikes in the light curve caused by cosmic ray contamination. The only case of a short GRB with ${\cal B}_0^1 > 1$ is GRB~171011A, with ${\cal B}_0^1 \approx 180$. The moderate Bayes factor in this case was caused by a signal that was identified with the interval between two spikes caused by cosmic rays. By limiting the maximum number of counts in one $100\, \mu{\rm s}$ bin to 8, we remove most of the contamination by cosmic rays from both short and long bursts (including 16 additional long burst outliers with $\log_{10}{\cal B}^1_0 > 10$, not shown).  After the removal of the cosmic ray contamination, the single long GRB outlier (dark gray)  with Bayes factor $\sim 6000$ is GRB 191004B; the signal was caused by 1000 Hz calibration pulses on the BAT detectors at nearly 300 s after the trigger, when Swift was slewing.}
\label{fig:B_Swift}
\end{figure}

Another false signal, which occurred in the Swift/BAT data for the long burst GRB~191004B, was caused by 1000-Hz pulses sent to BAT to aid in its calibration (D. Palmer, private communication; see the gray distribution in Extended Data Figure~\ref{fig:B_Swift}).

A third type of false signal can come from extraction of the data sets themselves. Kruger et al.  \cite{2002ApJ...576..932K} found in their search 
 a few cases (they highlight BATSE trigger 2101) in which there appeared to be significant power.  However, they discovered that in these cases the apparent signals were found in portions of the ASCII TTE data with an abnormally low count rate, which was not consistent with the original FITS data, and concluded that there had been corruption of the data sets when they were translated to ASCII.  Neither of our featured bursts have data corrupted in this way.

\begin{figure}[h!]
\centering
\includegraphics[width=0.45\textwidth]{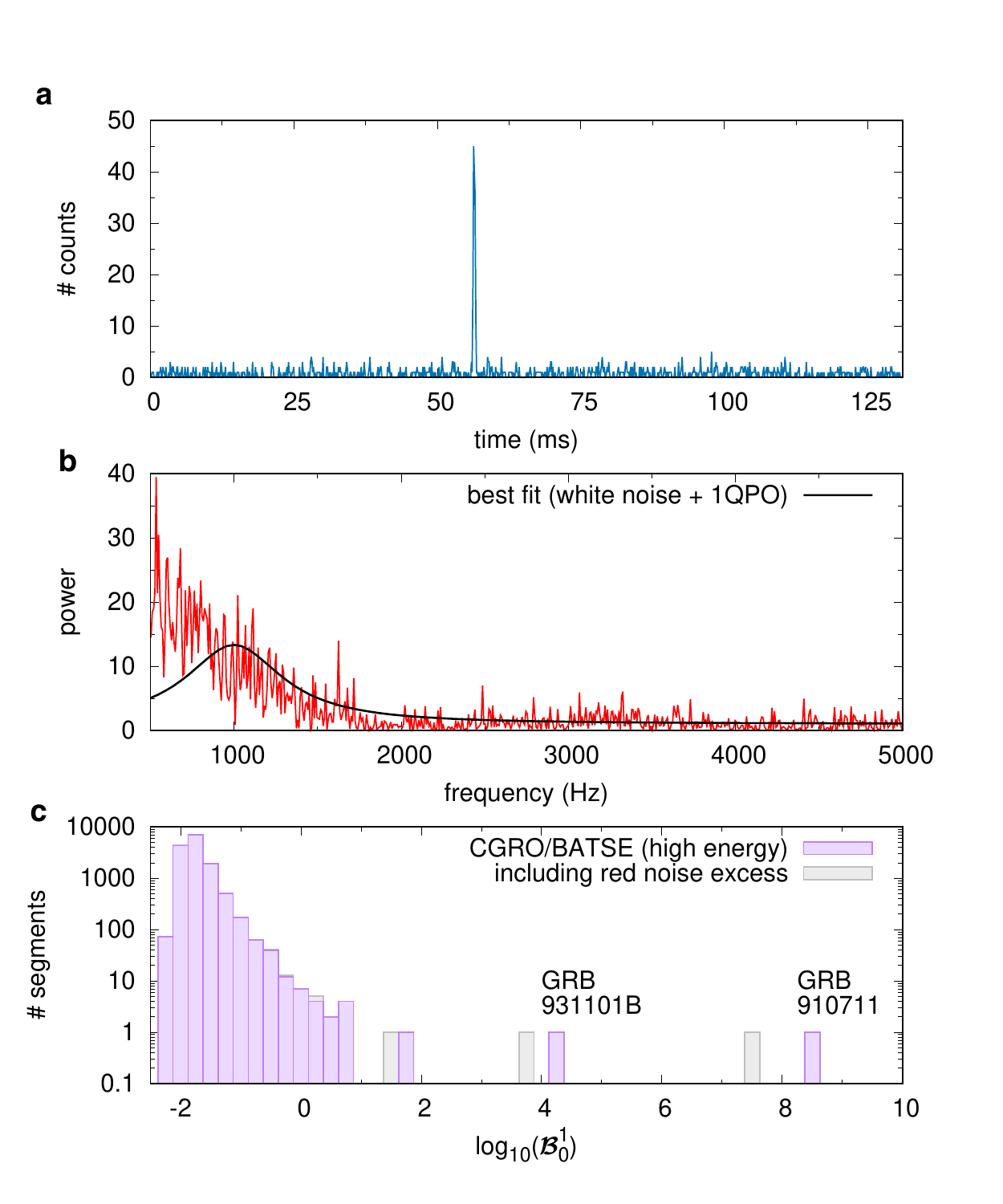}
\caption{$\vert$ \textbf{Analysis of BATSE segments contaminated by excess red noise.} \textbf{a,} Light curve of a segment near the middle of the $\sim 1$ second long burst GRB 980310B.  \textbf{b,} Power spectrum of that segment, which shows clear red noise above 500~Hz, and the best fit according to our algorithm.  The resulting Bayes factor for one QPO versus none is ${\cal B}_0^1\approx 10^{214}$.  \textbf{c,} The differential distribution of $\log_{10} {\cal B}^1_0$ (defined in the main text) for the entire set of BATSE short GRBs (there are 4 additional outliers with large amounts of red noise and $\log_{10} {\cal B}^1_0 > 9$, including the one featured in panels \textbf{a} and \textbf{b}), and for the subset obtained after removing data segments contaminated with red noise above 500 Hz.}
\label{fig:red_example}
\end{figure}

A fourth potential warning sign is the presence of large amounts of red noise up to hundreds of Hertz.  As indicated above we focus on frequencies above 500~Hz in large part to avoid the red noise which complicates our analysis.  But in a small number of cases the red noise is prominent even above 500~Hz (see panels \textbf{a} and \textbf{b} of Extended Data Figure~\ref{fig:red_example} for an example).  This makes it considerably more difficult to interpret excesses in the power density spectrum because, for example, it is not clear that the red noise should be a power law with a single slope.  A signature of red noise in our analysis, particularly of the BATSE data, is that the centroid frequency of a one-QPO fit is driven to the lowest allowed value, i.e., 1000~Hz. We use this signature to identify data segments with excess red noise. It is unclear what causes the large red noise in some of the BATSE bursts, but to avoid contamination we exclude these from our sample; see the bottom panel of Extended Data Figure~\ref{fig:red_example}. 

\begin{figure}[h!]
\centering
\includegraphics[width=0.45\textwidth]{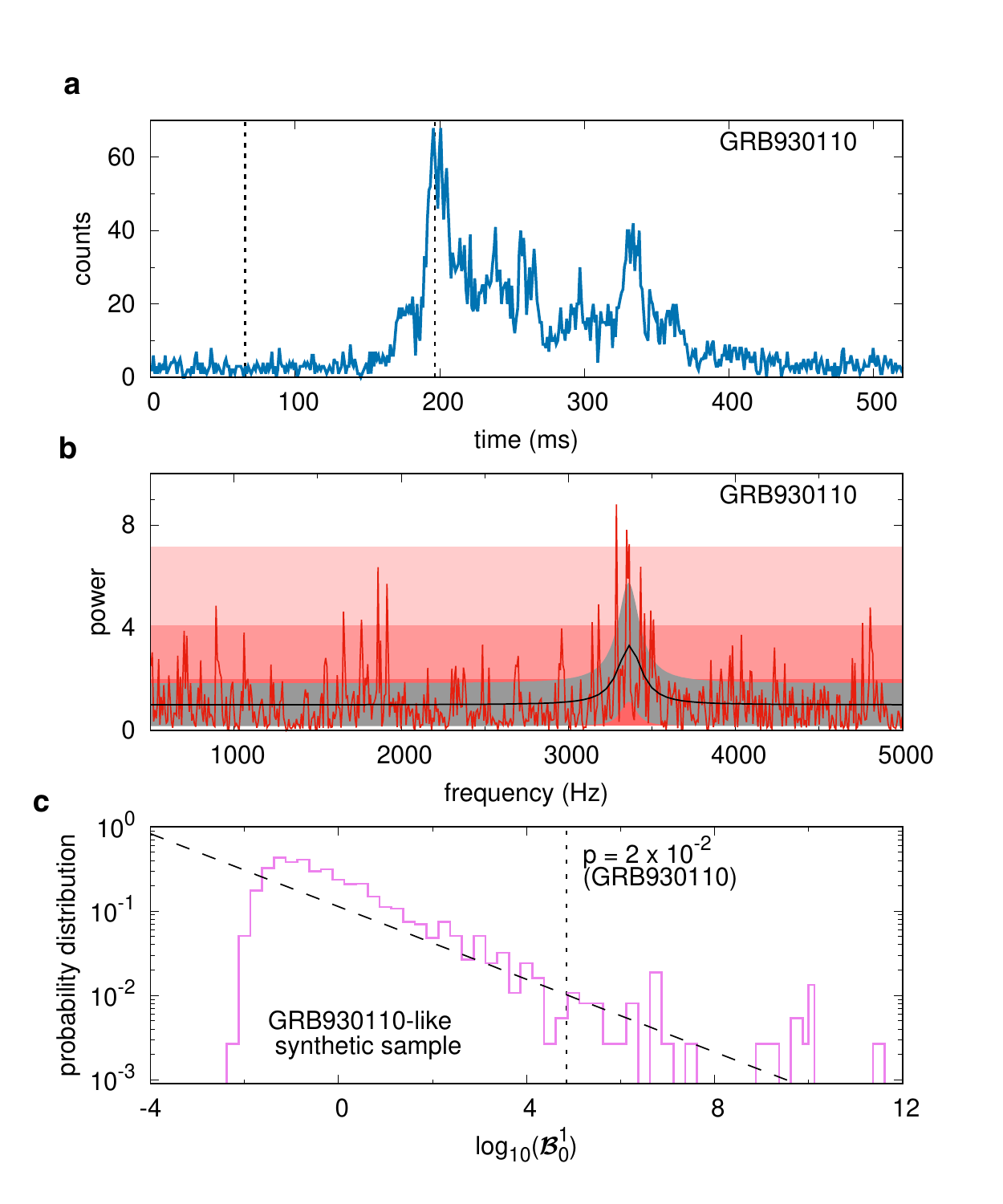}
\caption{$\vert$ \textbf{Example of a probably false signal of a QPO.}   \textbf{a,} BATSE data from GRB~930110, where the segment showing the apparent signal is framed by the two vertical dotted lines. Although GRB~930110 has $T_{90} = 220\,{\rm ms}$ \cite{1999ApJS..122..465P}, this segment is effectively an artificially very short GRB with $T_{90} \sim 10\, {\rm ms}$. \textbf{b,} Power spectrum for this segment and the best one-QPO plus white noise fit, along with the $1\sigma$, $2\sigma$, and $3\sigma$ power levels in the white noise only fit (cf. Figure~\ref{fig:powerspect}). \textbf{c,} Probability distribution of $\log_{10}{\cal B}_0^1$ generated with 1,500 realizations of light curves (without QPOs) fitted to the data segment shown in the top panel.  The vertical dotted line in the bottom panel shows the value of $\log_{10}({\cal B}_0^1)$ seen in the BATSE data for this segment of GRB~930110, the diagonal dashed line shows an exponential fit to the top 3\% of the Bayes factors, and the inset number shows the estimated false positive probability from the exponential fit. See the ``Generating synthetic data" and ``Estimates of significance of signals" sections below for more details.  Based on the high probability of a false positive (note also that in this case the next half-overlapping data segment has ${\cal B}_0^1 = 0.02$), we remove this segment from our sample. 12 other segments that presented similarly cropped light curves were also removed from our sample for consistency, although all cases had unremarkable Bayes factors.}
\label{fig:2125}
\end{figure}

Finally, a fifth potential cause for overestimating the significance of a QPO has been discussed recently \cite{2022ApJS..259...32H}. The basic effect is that if the total counts in a data segment are dominated by a small fraction of the segment, then a Fourier analysis of the whole segment effectively overresolves the power spectrum of the small contributing fraction.  As a result, contrary to the assumptions of such analyses, the power is correlated between frequencies and false positive QPOs can appear. See Extended Data Figure~\ref{fig:2125} for a discussion of one such case in the BATSE data we analyzed, probably caused by chance by the position of the edges of the data segment relative to the light curve of GRB~930110.

We combed the data for similar segments which included only the several milliseconds at the beginning or end of a burst, and removed all such segments from our sample.  None of those segments were as extreme as the first GRB~930110 segment, and indeed none of them had ${\cal B}_0^1>0.4$, but by removing them from the sample we decrease the probability of a false positive.

\bmhead{Test of goodness of fit for the QPOs}
As a crude test of goodness of fit, we can compute $\chi^2$ for just the frequencies of the QPO, in the range of the centroid plus or minus twice the width of the QPO, as suggested by H\"ubner et al.  \cite{2022ApJS..259...32H}.  The point is that when there are many frequencies in the power spectrum and only a small fraction of them are near a putative QPO, the overall $\chi^2$ can appear to be good even if the model is actually poor (or unreasonably good, as can happen if the power spectrum is oversampled) for the QPOs, because at most frequencies there is no excess power and the fit is acceptable.  To carry out this calculation we again use the Groth distributions \cite{1975ApJS...29..285G}, for which, given an expected (non-Poisson) power of $P_s$ in a single frequency bin, the mean power after including Poisson noise is $\langle P\rangle=1+P_s$ and the variance is $\langle P^2-\langle P\rangle^2\rangle=1+2P_s$.  Although the predicted power distribution is not Gaussian and thus a $\chi^2$ description is not strictly valid, this gives a rough indication of the goodness of fit.

Using this description, and including the three parameters per QPO when computing the number of degrees of freedom, we find that for the lower-frequency QPO in GRB~910711, $\chi^2/{\rm dof}=9.1/11$; for the higher-frequency QPO in GRB~910711, $\chi^2/{\rm dof}=11.6/12$; 
and the QPOs in GRB~931101B are too narrow, and thus have too small a number of degrees of freedom, for this to be a meaningful test.  As expected from good fits, both $\chi^2$ values are close to their respective numbers of degrees of freedom.

\begin{figure}[h!]
\centering
\includegraphics[width=0.45\textwidth]{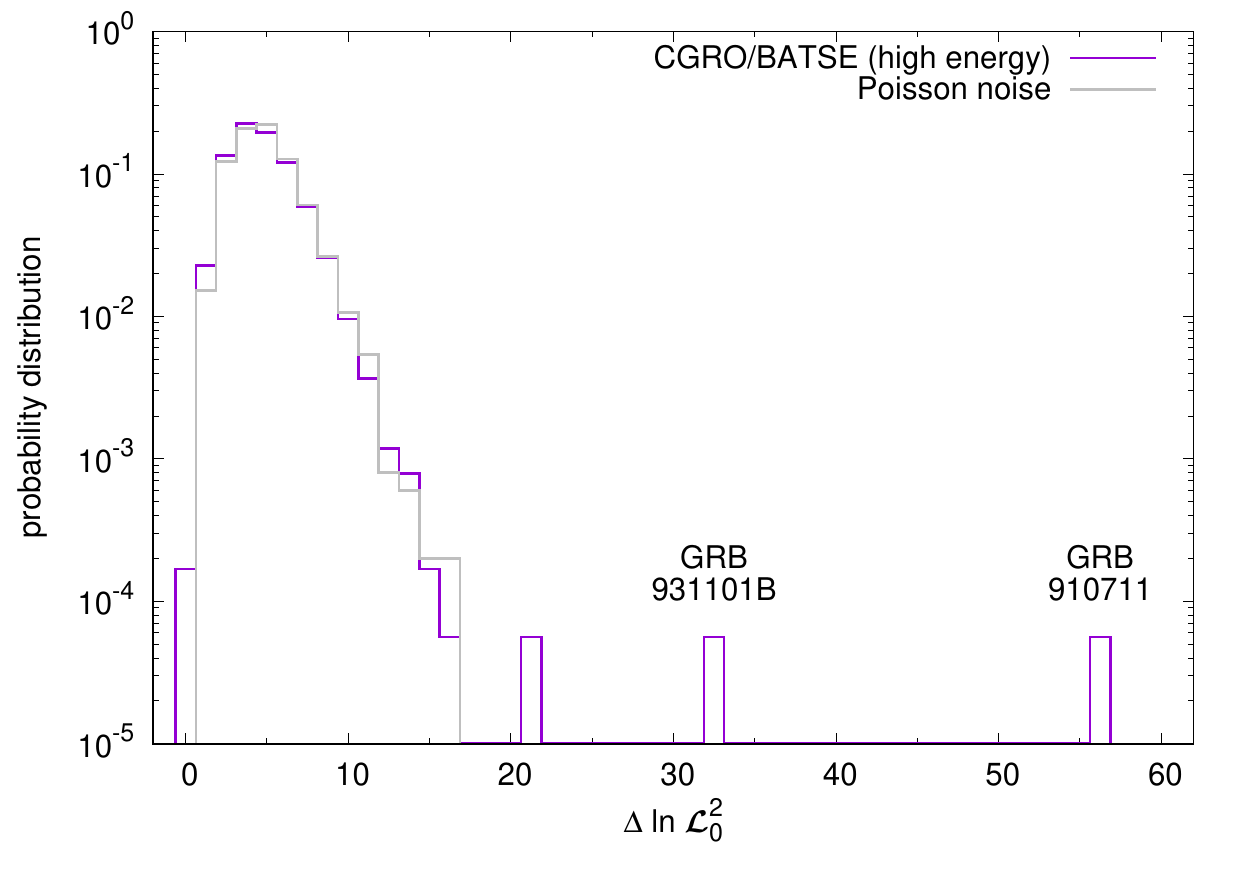}
\caption{$\vert$ \textbf{Probability distribution of signal strengths} Here we show the distribution of $\Delta\ln{\cal L}^2_0$ for the BATSE sample of short GRBs analyzed in our work, compared with the probability distribution obtained for 15,000 synthetic spectra generated containing independent realizations of Poisson noise. The two outliers are our signals, with $\Delta\ln{\cal L}^2_0 = 56.4$ (GRB~910711) and 33.3 (GRB~931101B). A third outlier with lower significance can also be seen at $\Delta\ln{\cal L}^2_0 = 21.3$. The bulk of the BATSE distribution (excluding the outliers) is well represented by the Poisson noise distribution.}
\label{fig:poisson}
\end{figure}

\bmhead{Distributions of $\Delta\ln{\cal L}^2_0$}
\label{sec:distribution}
From now on we focus only on the BATSE data, where our signals were detected. Because of the extra parameter volume required for a model with two Lorentzian QPOs plus white noise compared with a model that has zero or one QPO, it is not computationally feasible to calculate ${\cal B}^2_0$ for every segment of BATSE data (there are approximately 14,200 half-overlapping data segments in our analysis of 532 BATSE short bursts).  We can, however, get a sense for the evidence for two QPOs by looking at the distribution of $\Delta\ln{\cal L}^2_0$ , i.e. the difference between the maximum log likelihoods in a two-QPO model, and the maximum log likelihood in a white noise only model; this will be our figure of merit in the Section ``Estimates of significance of signals'' below. (The calculation of the significance using the difference between maximum log likelihoods is less robust than using the Bayes factor, as it does not take into account the structure of the likelihood surface. However, a test comparing $\Delta \ln{\cal L}^1_0$ and ${\cal B}^1_0$ shows that they give very similar results.) In Extended Data Figure~\ref{fig:poisson} we show the probability distribution of $\Delta\ln{\cal L}^2_0$ inferred for the BATSE sample and compare it with a distribution obtained from synthetic realizations of Poisson noise. The bulk of the BATSE distribution is well represented by the Poisson noise distribution.

\begin{figure*}[h!]
\centering
\includegraphics[width=1.0\textwidth]{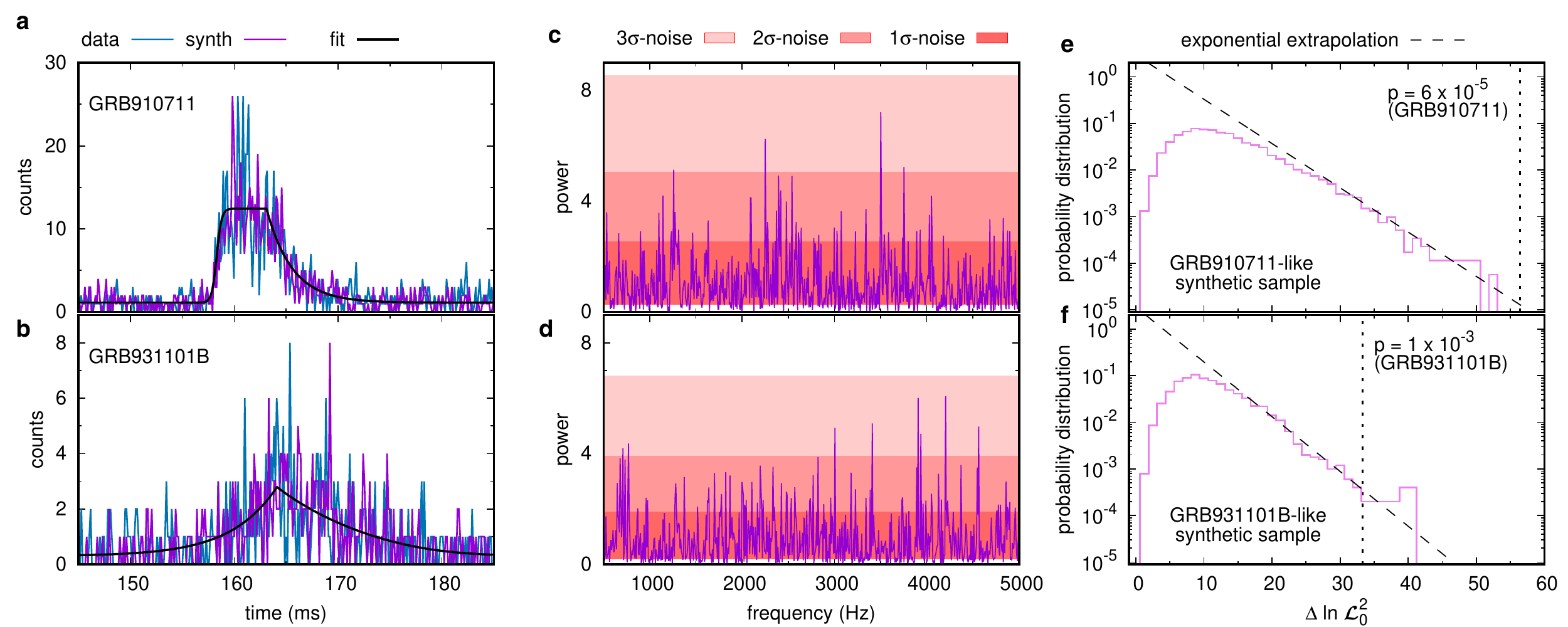}
\caption{$\vert$ \textbf{Real versus synthetic data} \textbf{a,} Zoom-in on the QPO data segment for GRB~910711, a corresponding smoothed fits given by eq. (\ref{eq:synthmodel}) and a representative example of the synthetic light curve obtained via Poisson sampling from the smoothed fits (the starting time of the GRB~910711 light curve is shifted here for convenience of presentation). \textbf{c,} Power spectrum of the synthetic light curve shown in \textbf{a}. As in Figure~\ref{fig:powerspect}, the red bands show the $1\sigma$, $2\sigma$, and $3\sigma$ powers expected given the best white noise only fits to the data from each burst.  \textbf{e,} Probability distribution of $\Delta\ln{\cal L}_0^2$ from synthetic data generated from light curve fits to GRB~910711.  The vertical dotted line shows the $\Delta\ln{\cal L}_0^2$ observed in BATSE data, and the diagonal dashed line shows the exponential fit to the top 3\% of the synthetic data points (see ``Estimates of significance of signals" for details).  The inset number gives the estimated false positive probability for signals as strong as or stronger than that observed.  \textbf{b, d, f,} Similarly, for GRB~931101B.} 
\label{fig:synth_data}
\end{figure*}

\bmhead{Generating synthetic data}
\label{sec:synthetic_data}
When a data stream is not stationary, as it is not in our case because the bursts present rapidly-changing flux, then power density spectra constructed using fast Fourier transforms can lead to overestimates of the significance of QPOs \cite{2022ApJS..259...32H}.  To assess the impact of this concern in our case, we generate synthetic data with the procedure described below, following H\"ubner et al. \cite{2022ApJS..259...32H}:

\begin{enumerate}
\item We model (without QPOs) the light curve of each data segment in which we find candidate signals.  The signal model we use for the count rate as a function of time is related to a hyperbolic tangent and has six parameters as a function of the time $t$:
\begin{equation}
F=F_{\rm back}+F_m\{1+\tanh[(t-t_m)/t_r]\}
\label{eq:synthmodel}
\end{equation}
times 1 if $t<t_p$ or times $\exp[-(t-t_p)/t_d]$ if $t\geq t_p$.  Thus $F_{\rm back}$ is the background count rate, $F_m$ is related to the flux of the burst, $t_m$ is a characteristic time for the onset of the burst, $t_r$ is the rise time, $t_p$ is the time when the exponential decay starts, and $t_d$ is the decay time.  Other fits are possible.  For example, we could use a fit in which the light curve is represented by one exponential rising to a peak, and then another exponential falling from that peak \cite{2022ApJS..259...32H}.  However, we find that our fit is preferred strongly: for example, for GRB~910711, the maximum likelihood of our fit is $\sim 8\times 10^5$ times larger than the maximum likelihood of the double-exponential fit.

In Extended Data Table~\ref{tab:synthparams} we give the best-fit values of these six parameters for each of the two segments in which we find signals.  Note that for GRB~931101B, $t_p<t_m$, which simply indicates that this burst has a somewhat gradual rise and decline.

\begin{table*}[h!]
\begin{center}
\begin{minipage}{\textwidth}
\caption{$\vert$ \textbf{Model parameters used in synthetic data}}\label{tab:synthparams}
\setlength{\tabcolsep}{5pt} 
\begin{tabular}{lcccccc}
\toprule
GRB & $F_{\rm back}$  & $F_m$ & $t_m$ & $t_r$ & $t_p$  & $t_d$  \\
\midrule
910711  & 8.764 & 44.209 & 98.323 & 0.472 & 103.11 & 2.472 \\
931101B & 2.330 & 255.65 & 46.950 & 8.609 & 33.048 & 2.958  \\
\bottomrule
\end{tabular}
\footnotetext{Note: parameters for our synthetic data runs modeling QPO-free light curves based on models of the light curves from our two BATSE data segments where we find QPOs.  See text and especially Equation~\ref{eq:synthmodel} for details.  Here the count rates $F_{\rm back}$ and $F_m$ are in counts per millisecond, and the times are all in milliseconds; the zeros of time for $t_m$ and $t_p$ are shown in Figure~\ref{fig:lightcurves}. }
\end{minipage}
\end{center}
\end{table*}

\item We perform a Poisson draw at each time bin of 2 microseconds to obtain a light curve.  These light curves, by design, have no QPOs and thus analysis of them gives a sense for the probability of a false positive identification of a QPO or QPOs given the non-stationary count rate.  In panels \textbf{a} and \textbf{b} of Extended Data Figure~\ref{fig:synth_data} we show the data, the smoothed fit, and a representative Poisson draw for each of our two bursts.

\item We then compute the power spectrum and are ready to perform our analysis to search for QPOs in the synthetic data. 
In panels \textbf{c} and \textbf{d} of Extended Data Figure~\ref{fig:synth_data} we show the power spectra obtained for the representative examples presented in the left panel.
\end{enumerate}

\bmhead{Estimates of significance of signals}
We perform the steps outlined in the previous Section repeatedly for each of our two bursts to search the synthetic data for signals as strong as or stronger than those we found in the BATSE data.  This gives us two distributions of $\Delta\ln{\cal L}_0^2$, representative of GRB~910711-like and GRB~931101B-like light curves. For each burst, we therefore conclude:

\begin{enumerate}
\item For GRB~910711-like synthetic data, we perform $\approx 14,000$ simulations and find none with $\Delta\ln{\cal L}_0^2$ as large as in the actual GRB~910711 data, which suggests that the false positive probability for this signal is $\ltorder 10^{-4}$.

\item For GRB~931101B-like synthetic data, we perform 4,000 simulations and find 7 with $\Delta\ln{\cal L}_0^2$ at least as large as in the actual GRB~931101B.  Thus we can roughly assign a probability of $\sim 1-2\times 10^{-3}$ for the GRB~931101B signal.
 \end{enumerate}

We show the distributions of the synthetic $\Delta\ln{\cal L}_0^2$ for GRB~910711-like and GRB~931101B-like data in panels \textbf{e} and \textbf{f} of Extended Data Figure~\ref{fig:synth_data}.  Note that these distributions extend to far larger $\Delta\ln{\cal L}$ than the bulk of our distribution of BATSE data (see Extended Data Figure~\ref{fig:poisson}); in this sense, GRB~910711 and GRB~931101B are not typical bursts.

Given that none of our $\approx 14,000$ GRB~910711-like synthetic data sets reached a $\Delta\ln{\cal L}_0^2$ as high or higher than what we see in the BATSE data, to obtain a more precise estimate of the significance of this signal we need to extrapolate.  We see from Extended Data Figure~\ref{fig:synth_data} that at the high-$\Delta\ln{\cal L}_0^2$ end, the distribution appears linear on a log-linear plot, and is therefore well fit with an exponential.  Performing such a fit using the top 3\% of $\Delta\ln{\cal L}_0^2$ values for the GRB~910711-like synthetic data sets yields a probability, in a single realization, of
\begin{equation}
\label{eq:prob_exp}
{\rm Prob}(\Delta\ln{\cal L}_0^2 \geq x)=
0.03\ e^{-\frac{(27.9-x)}{4.57}}. 
\end{equation}
For $\Delta\ln{\cal L}_0^2>56.4$ (the value for GRB~910711), this implies ${\rm Prob}(\Delta\ln{\cal L}_0^2 > 56.4) \approx 6\times 10^{-5}$. That is, the probability of a false positive, \textit{in a single realization}, for the QPOs in GRB~910711 is approximately $6\times 10^{-5}$.

We now need to estimate the probability of $\Delta\ln{\cal L}_0^2>56.4$ taking into account the number of trials, that is, considering \textit{all} of the bursts that we analyze (as the same method is used in both cases, the search for QPOs in each synthetic light curve has the same number of trials as the search for QPOs in each BATSE light curve).  We find that, for each burst, the probability of a given large $\Delta\ln{\cal L}_0^2$ depends strongly on the duration $T_{90}$ of the burst (although obviously the specific shape of the burst light curve and the number of counts in the burst also play roles).  This is plausible given the argument that the probability of a false positive in a segment is greater when the duration of the burst is a smaller fraction of the duration of the segment \cite{2022ApJS..259...32H}.
Quantitatively, we performed 2,000 simulations each on the light curves of five other $T_{90}<100$~ms BATSE bursts with good data \cite{1999ApJ...527..827C}: these were GRBs 910508, 910625, 910703, 930113C, and 940621C.  As with GRB~910711, we performed exponential fits to the top 3\% of $\Delta\ln{\cal L}_0^2$ values.  In Extended Data Table~\ref{tab:extrapprob} we show the extrapolated probability of $\Delta\ln{\cal L}_0^2>56.4$ for each of these bursts, along with their $T_{90}$.

\begin{table*}[h!]
\begin{center}
\begin{minipage}{\textwidth}
\caption{$\vert$ \textbf{Extrapolated probabilities of $\Delta\ln{\cal L}_0^2$ for different short bursts}}\label{tab:extrapprob}
\setlength{\tabcolsep}{4pt} 
\begin{tabular}{lccccc}
\toprule
GRB & Trigger \# & $T_{90}$ (ms) & Counts & ${\rm Prob}(\Delta\ln{\cal L}_0^2>56.4)$ & ${\rm Prob}(\Delta\ln{\cal L}_0^2>33.3)$\\
\midrule
{\textbf{910711}} & 512  & 14 & 1790 & $5.9\times 10^{-5}$ & $9.2\times 10^{-3}$\\
910508 & 207  & 30 & 1254 & $2.2\times 10^{-6}$ & $1.6\times 10^{-3}$\\
{\textbf{931101B}} & 2615 & 34 & 524 & $2.6\times 10^{-6}$ & $1.3\times 10^{-3}$\\
910625 & 432  & 50 & 1810 & $7.2\times 10^{-7}$ & $9.3\times 10^{-4}$\\
910703 & 480  & 62 & 2278 & $1.8\times 10^{-7}$ & $7.5\times 10^{-4}$\\
940621C & 3037 & 66 & 710 & $2.0\times 10^{-10}$ & $7.9\times 10^{-6}$\\
930113C & 2132 & 90 & 612 & $4.1\times 10^{-11}$ & $2.9\times 10^{-6}$\\
\bottomrule
\end{tabular}
\footnotetext{Note: probabilities, extrapolated using an exponential fit, that synthetic data generated using the light curves of each of the bursts with durations $T_{90}<100$~ms (third column) and good data would produce $\Delta\ln{\cal L}_0^2>56.4$ (fifth column), which is the value obtained using the BATSE data for GRB~910711, and $\Delta\ln{\cal L}_0^2>33.3$ (sixth column), which is the value obtained using the BATSE data for GRB~931101B.  The fourth column gives the total number of counts in the 0.131072-second segment that we analyzed, rather than the counts summed over the $T_{90}$ of the burst.  We see that both probabilities decrease steeply with increasing duration. The bursts where we find QPOs are highlighted in boldface.}
\end{minipage}
\end{center}
\end{table*}

We see in Extended Data Table~\ref{tab:extrapprob} that there is a steep decline in the probability of a high $\Delta\ln{\cal L}_0^2$ with increasing $T_{90}$.  Based on this analysis, it seems likely that the expected number of false positives with $\Delta\ln{\cal L}_0^2>56.4$ in the \textit{entire} sample of bursts is dominated by the expected number of false positives for the \textit{single} burst GRB~910711 (as the probabilities in Extended Data Table~\ref{tab:extrapprob} are small [given by exponential extrapolations of the form (\ref{eq:prob_exp})], they can be used interchangeably with the number of false positives in each case).  That is, the expected number of false positives in the entire catalog, $N_{\rm false,catalog}(\Delta\ln{\cal L}_0^2>56.4)$, is
\begin{eqnarray*}
     \sum_i N_{\rm false,i}\left(\Delta\ln{\cal L}_0^2>56.4\right) \sim 6\times 10^{-5}\,,
\end{eqnarray*}
where $i$ runs over all bursts. We therefore conclude that the significance of the QPO signal in GRB~910711, when all trials over all bursts are taken into account, is $\sim 6\times 10^{-5}$.

We can then ask the question: what is the probability that in our sample we would see a burst with $\Delta\ln{\cal L}_0^2>56.4$ (like GRB~910711) and a second burst with $\Delta\ln{\cal L}_0^2>33.3$ (like GRB~931101B)?  In Extended Data Table~\ref{tab:extrapprob} we therefore also show the extrapolated probability of $\Delta\ln{\cal L}_0^2>33.3$.  The expected number of false signals in the whole catalog is then $N_{\rm false,catalog}(\Delta\ln{\cal L}_0^2>56.4\ {\rm and}\ \Delta\ln{\cal L}_0^2>33.3)$, given by
\begin{eqnarray*}
\sum_i \Biggl\{N_{\rm false,i}\left(\Delta\ln{\cal L}_0^2>56.4\right) \times \\
\left[\sum_{i\neq j} N_{\rm false,i}\left(\Delta\ln{\cal L}_0^2>33.3\right)\right]\Biggr\} \sim 3\times 10^{-7}\,,
\end{eqnarray*}
where $i$ again runs over all bursts and $j$ runs over all bursts other than $i$.  We thus conclude that the combined significance of the QPO signals detected in both GRB~910711 and GRB~931101B is $\sim 3\times 10^{-7}$, taking into account all trials over all bursts. The addition of the Swift/BAT and Fermi/GBM bursts we analyzed does not impact this result. There were 14 Fermi/GBM bursts with $T_{90} < 100\,{\rm ms}$, but the GBM data included only the duration of each burst in each case (not including portions of pre- and post-burst low-count background data), therefore mitigating this possible cause of false QPOs \cite{2022ApJS..259...32H}. There were no Swift/BAT bursts in our sample with $T_{90} < 100\,{\rm ms}$.

\bmhead{Short GRBs vs. SGR giant flares}

As noted in the main text, tentative evidence for a kilohertz QPO in a giant flare from the Galactic SGR~1806$-$20 has been reported \cite{2006ApJ...653..593S} as well as kilohertz QPOs from an SGR giant flare in NGC~253 \cite{2021Natur.589..207R}.  Moreover, it has been proposed that many of the shortest gamma-ray bursts are giant flares from SGRs rather than neutron star mergers \cite{2021ApJ...907L..28B}.  It is therefore useful to explore how to distinguish the giant flare from the neutron star merger scenario.

The most definitive distinction would come from an identification of the host galaxy, because SGR giant flares do not reach the isotropic equivalent energy or luminosity of double neutron star mergers.  However, BATSE localizations are only to several square degrees and thus clear host identification is impossible.

Nonetheless, energetics do distinguish between our bursts and the giant flare sample \cite{2021ApJ...907L..28B}.  Extended Data Table~\ref{tab:flux} gives the implied minimum equivalent isotropic energy and isotropic peak luminosity for each of our bursts, based on the minimum distance of a galaxy consistent with the burst localizations.  We see that even if we take the minimum galaxy distance irrespective of star formation rate, the required $L_{\rm iso}$ are extreme; for comparison, the maximum $L_{\rm iso}$ in the candidate giant flare list \cite{2021ApJ...907L..28B} was $1.8\times 10^{48}$~erg~s$^{-1}$.

If we restrict consideration to only galaxies with at least moderately active star formation, given that magnetar giant flare activity is thought to come from neutron stars with very young ages $\ltorder 10^4$~years, then both $E_{\rm iso}$ (compare with the maximum of $5.3\times 10^{46}$~erg in the giant flare sample \cite{2021ApJ...907L..28B}) and $L_{\rm iso}$ stand out compared with the suggested giant flare sample.  This provides an argument that our bursts are classical short gamma-ray bursts, and thus are likely to be produced by neutron star mergers, rather than being giant flares from SGRs.

\begin{table*}[h!]
\begin{center}
\begin{minipage}{\textwidth}
\caption{$\vert$ \textbf{Fluences, fluxes, and distances of bursts with QPOs}}\label{tab:flux}
\setlength{\tabcolsep}{3pt} 
\begin{tabular}{lcc}
\toprule
\multicolumn{1}{r}{GRB} &910711 & 931101B \\
\midrule
$>20$~keV fluence (erg~cm$^{-2}$) & $4.3\times 10^{-7}$ & $1.8\times 10^{-7}$\\
Estimated maximum flux (erg~cm$^{-2}$~s$^{-1}$) & $1.5\times 10^{-4}$ & $2.6\times 10^{-5}$\\
Minimum distance of galaxy (Mpc) & 15.6 & 24.2\\
Minimum $E_{\rm iso}$ (erg) & $1.2\times 10^{46}$ & $1.2\times 10^{46}$ \\
Minimum peak $L_{\rm iso}$ (erg~s$^{-1}$) & $5\times 10^{48}$ & $2\times 10^{48}$ \\
Minimum distance of star forming galaxy (Mpc) & 66.3 & 46.4\\
Minimum $E_{\rm iso}$ for star forming galaxy (erg) & $2.3\times 10^{47}$ & $4.8\times 10^{46}$\\
Minimum peak $L_{\rm iso}$ for star forming galaxy (erg~s$^{-1}$) & $9\times 10^{49}$ & $7\times 10^{48}$\\
\bottomrule
\end{tabular}
\footnotetext{Note: here $E_{\rm iso}$ and $L_{\rm iso}$ are respectively the equivalent isotropic energy release and isotropic peak luminosity that would give the observed fluence and peak flux at the listed distances.  We use the GLADE+ sample \cite{2018MNRAS.479.2374D,2021arXiv211006184D}, along with GRB localizations\cite{1996ApJS..106...65M} (which is also our source for the $>20$~keV fluence), to determine the closest galaxy, and the closest galaxy consistent with the direction to the GRB that has an absolute B magnitude equal to or brighter than the $M_B=-20.8$ for the Galaxy \cite{2004AJ....127.2031K}.  The B magnitude threshold is based on the suggestion that giant flares from SGRs should be correlated with ongoing star formation because giant flares are thought to occur no more than $\sim 10^4$~years after the birth of a neutron star\cite{2021ApJ...907L..28B}.}
\end{minipage}
\end{center}
\end{table*}

\bmhead{Comparison with numerical relativity results}
\label{sec:compareNR}
The frequencies of the main QPO peak in simulations of binary neutron star mergers range from $1.8 - 3.8$ kHz \cite{2014PhRvL.113i1104T}, depending on the neutron star equation of state and on the individual masses of the neutron stars in the binary. The values obtained for these frequencies are remarkably consistent across different groups, as evidenced by tests which achieved differences smaller than the scatter of the proposed phenomenological relations \cite{2017ApJ...842L..10R}. These frequency values are also consistent with the QPOs we present in this work. The quality factors of the $\nu_2$ QPOs we find are $Q \sim 50 - 100$, which are a few times times higher than the quality factors $Q \sim 10 - 20$ estimated from the simulations \cite{2005PhRvL..94t1101S,2012PhRvL.108a1101B,2014PhRvL.113i1104T,2017ApJ...842L..10R}; in contrast to the frequencies, which are very similar among different simulations, there is less agreement between simulations about the quality factor. This is because the lifetime of an HMNS in a simulation is sensitive to numerical details of the evolution codes. Additionally, it is expected that the quality factors obtained from numerical relativity are lower limits, due to numerical dissipation \cite{2019PhRvD..99h4024R}.

\bmhead{Additional analysis with adjustable red noise}
We have also performed analyses of our two signals using red noise instead of white noise.  For these analyses, the noise contributes $A_{\rm red}(f/500~{\rm Hz})^p$ to the power, where the prior for $A_{\rm red}$ is flat between 0 and 5 and the prior for $p$ is flat between $-3$ and $+1$ (thus allowing for blue noise).

From Extended Data Table ~\ref{tab:rednoise} we see that the log likelihoods of the noise only, noise + 1 QPO and noise + 2 QPO fits are not increased significantly when, in each case, white noise is replace by red noise. Therefore, the fits are not improved significantly by the use of red noise instead of white noise. 

Moreover, Extended Data Table ~\ref{tab:rednoise} shows that $\Delta \ln {\cal L}^2_0$ for GRB~910711 is only slightly enhanced if we use the red noise models ($\Delta \ln {\cal L}^2_0 = 57.2-0.1 = 57.1$) compared with the white noise models ($\Delta \ln {\cal L}^2_0 = 56.4$); for GRB~931101B, $\Delta \ln {\cal L}^2_0$ is slightly reduced if we use the red noise models ($\Delta \ln {\cal L}^2_0 = 33.6-2.3 = 31.3$) compared with the white noise models ($\Delta \ln {\cal L}^2_0 = 33.3$). From this we conclude that the choice of noise model (red or white) does not affect how much the noise + 2 QPOs model is preferred over the noise-only model.  Finally, Extended Data Table ~\ref{tab:rednoise} shows that the best-fit frequencies and frequency widths are virtually identical between the white noise and the red noise fits.

\begin{table*}[h!]
\begin{center}
\begin{minipage}{\textwidth}
\caption{$\vert$ \textbf{Comparison between white noise and red noise fits}\label{tab:rednoise}}
\setlength{\tabcolsep}{1pt} 
\begin{tabular}{lccccccc}
\toprule
GRB & Fit   & Slope & $\nu_1$ (Hz) & $\Delta\nu_1$ (Hz) & $\nu_2$ (Hz) & $\Delta\nu_2$ (Hz)  & $\ln{\cal L}_{\rm best}-\ln{\cal L}_{\rm white}$  \\
\midrule
910711  & White & &  & &  &  & 0.0\\
              & Red & $-0.15_{-0.2}^{+0.19}$ & & & & & 0.1\\
              & White + 1QPO & & & & $2649_{-7}^{+6}$ & $26_{-7}^{+9}$ & 29.7\\
              & Red + 1QPO & $-0.28_{-0.23}^{+0.23}$ & & & $2649_{-7}^{+7}$ & $24_{-7}^{+9}$ & 30.0\\
              & White + 2QPOs & & $1113_{-8}^{+7}$ & $25_{-7}^{+9}$ & $2649_{-7}^{+6}$ & $26_{-7}^{+9}$ & 56.4\\
              & Red + 2QPOs & $+0.15_{-0.41}^{+0.36}$ & $1112_{-9}^{+8}$ & $27_{-7}^{+9}$ & $2648_{-8}^{+7}$ & $28_{-8}^{+10}$ & 57.2\\
931101B & White & &  &  &  &  &  0.0 \\
               & Red & $-2.00_{-0.61}^{+0.65}$ & & & & & 2.3\\
               & White + 1QPO & & & & $2612_{-8}^{+9}$ & $14_{-3}^{+7}$ & 20.5\\
               & Red + 1QPO & $-2.27_{-0.44}^{+0.56}$ & & & $2611_{-7}^{+5}$ & $14_{-3}^{+5}$ & 22.5\\
               & White + 2QPOs & & $877_{-8}^{+6}$ & $15_{-2}^{+7}$ & $2612_{-8}^{+9}$ & $14_{-3}^{+7}$ & 33.3\\
               & Red + 2QPOs & $-2.19_{-0.44}^{+0.74}$ & $879_{-10}^{+10}$ & $16_{-4}^{+12}$ & $2611_{-7}^{+9}$ & $15_{-4}^{+6}$ & 33.6\\
               \bottomrule
\end{tabular}
\footnotetext{Note: fits to the power densities of our two signals using our fiducial white noise background, and using instead a red noise background (see text for details).  In each case we show the median and the $\pm 1\sigma$ values of the parameters, and the final column gives the log likelihood of the best fit minus the log likelihood of the white noise only fit.  The ``Slope" column gives the slope of the noise fit when the noise is not required to be white; negative is red noise and positive is blue noise, and note that when red noise is preferred it only affects the lowest frequencies in our 500~Hz to 5000~Hz interval.  Using red instead of white noise does not change significantly the parameter values or the delta log likelihoods.}
\end{minipage}
\end{center}
\end{table*}

\bmhead{Use of Poisson only instead of white noise as background}
Our fiducial models include white noise as well as, possibly, Lorentzian QPOs.  As indicated above, the motivation for considering white noise is that spikes in the Swift/BAT data introduce large amounts of extra white noise into the power spectra, so this needs to be taken into account.  However, the BATSE data sets we analyze, after removal of segments with large amounts of $f>500$~Hz red noise, do not have spikes.  This suggests that, instead, it could be interesting to consider models in which the noise is purely Poisson, e.g., for a noise-only model the picture would be that the intrinsic count rate is steady and thus the only contributor to the power spectrum is photon counting noise, i.e., Poisson noise.

Although we have not reanalyzed all of the data sets with Poisson+QPO(s) models, a Poisson-only model is fast to evaluate.  When we do this, we find that our strongest signal GRB~910711 stands out even more from the synthetic light curves.  Exponential extrapolations of the type describe above suggest that the expected number of false positives per GRB~910711-like light curve is $\sim{\rm few}\times 10^{-7}$, i.e., that the signal is at least $100\times$ more significant than what we inferred from our white noise models.  (The signal would be even more significant if we were to make the [small] correction for detector deadtime, which lowers the Poisson power, as seen for example in the discovery of kHz QPOs in Scorpius X-1   \cite{1996ApJ...469L...1V}.)

As an aside, we also note that when we analyze the BATSE GRB~910711 data using models with Poisson noise plus some number of QPOs, this analysis identifies a third, weaker but still significant, QPO centered at $\sim 2070$~Hz with a width of $\sim 90$~Hz, and a fourth, even weaker and not obviously significant, QPO centered at $\sim 3700$~Hz with a width of $\sim 40$~Hz.  However, because we have not performed systematic studies using a Poisson noise background, we cannot assess these implications thoroughly.

\begin{figure}[h!]
\centering
\includegraphics[width=0.45\textwidth]{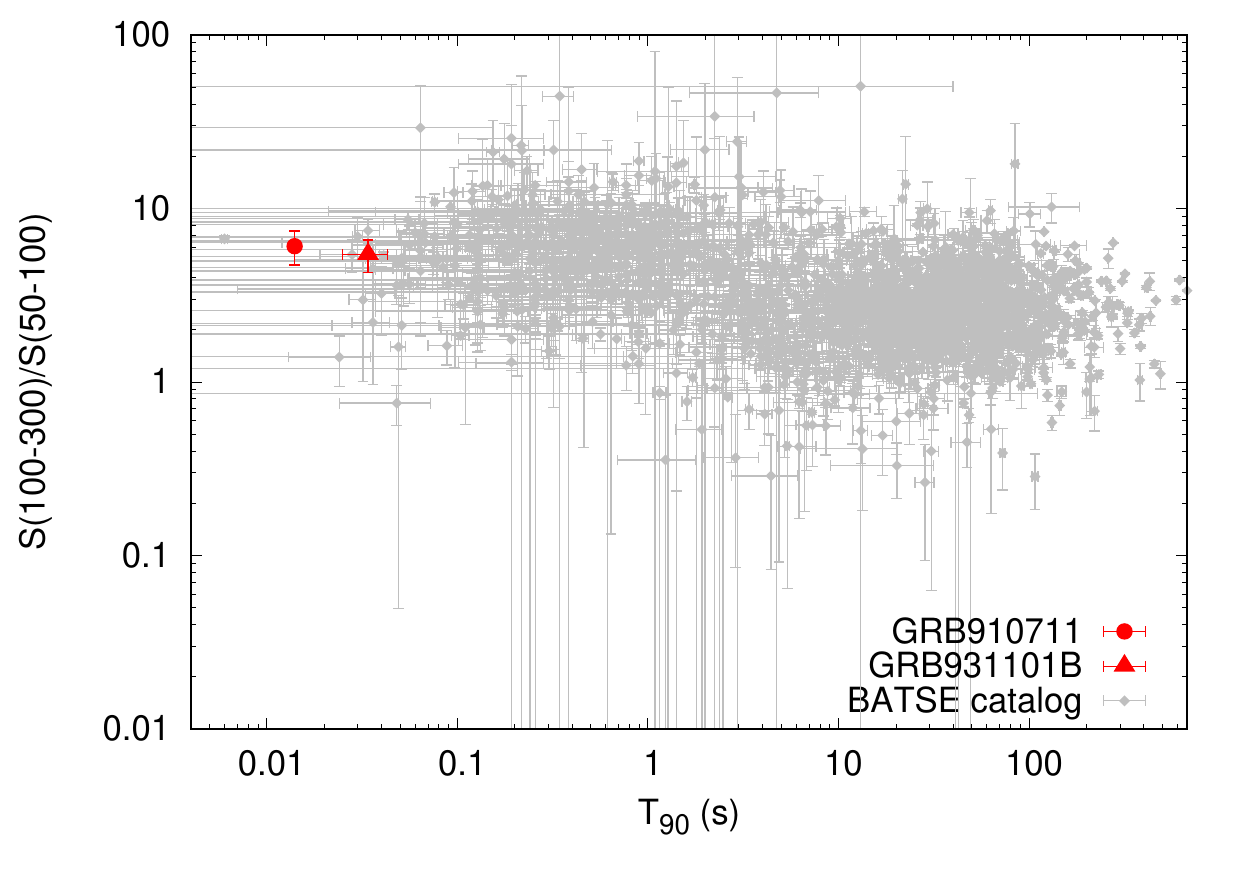}
\caption{$\vert$ \textbf{Hardness-duration plot of BATSE gamma-ray bursts} Here we show the hardness ratio (fluence in the $100-300$~keV band divided by fluence in the $50-100$~keV band) {\it vs.} $T_{90}$ for the BATSE catalog (including both short and long GRBs), highlighting our signals. Error bars represent $\pm 1\sigma$ uncertainties. The $T_{90}$ of some of the shortest bursts was re-calculated using the TTE data \cite{1999ApJ...527..827C}.  GRB~910711 and GRB~931101B are very short compared with most short GRBs,  but the hardness ratios of our bursts do not stand out in the short GRB population.}
\label{fig:HRxT90}
\end{figure}

\bmhead{Detailed analysis of candidate signals}
\label{sec:signals}
Extended Data Table~\ref{tab:bestfits} shows the best-fit parameters for our two-QPO models for each of our two bursts.  Extended Data Figure~\ref{fig:HRxT90} places our bursts on a hardness ratio-duration plot for all BATSE bursts.

\begin{table*}[h!]
\begin{center}
\begin{minipage}{\textwidth}
\caption{$\vert$ \textbf{Best two-QPO fit to data from each burst}}\label{tab:bestfits}
\setlength{\tabcolsep}{2.5pt} 
\begin{tabular}{lcccccccccc}
\toprule
GRB & Trigger \# & $A_{\rm white}$ & $A_1$ & $a^{\rm rms}_{{\rm osc}_1}$ & $\nu_1 ({\rm Hz})$ & $\Delta\nu_1$ (Hz) & $A_2$ & $a^{\rm rms}_{{\rm osc}_2}$ & $\nu_2$ (Hz) & $\Delta\nu_2$ (Hz) \\
\midrule
910711  & 512 & $0.18$ & $6.20$ & $0.27$ & $1113$ & $26$ & $7.10$ & $0.28$ & $2650$ & $28$\\
931101B & 2615 & $0.003$ & $5.57$ & $0.32$ & $878$ & $13$ & $7.55$ & $0.34$ & $2613$ & $12$ \\
\bottomrule
\end{tabular}
\footnotetext{Note: the parameter values for the best two-QPO fit to the power spectral data for each burst, which are also labeled by their BATSE trigger number.  Here we also add the best estimate of the fractional root mean squared amplitudes of each QPO. 
The best values of the parameters need not be the same as the median values from Table~\ref{tab:burstqpos}, but the best values are all within the $\pm 1\sigma$ ranges.  For GRB~931101B the best-fit white-noise amplitude is very low, which indicates that the model has captured the signal power.  In contrast, the best-fit white-noise amplitude for GRB~910711 is larger, which could suggest that there are additional QPOs or other features, but the evidence is not strong.  If, for example, we add a third QPO to our fit then the best frequency is $\sim 2100$~Hz and the best frequency width is $\sim 100$~Hz, but the evidence is only $\sim 50$\% larger than the two-QPO evidence.}
\end{minipage}
\end{center}
\end{table*}

\begin{figure*}[h!]
\centering
\includegraphics[width=1.0\textwidth]{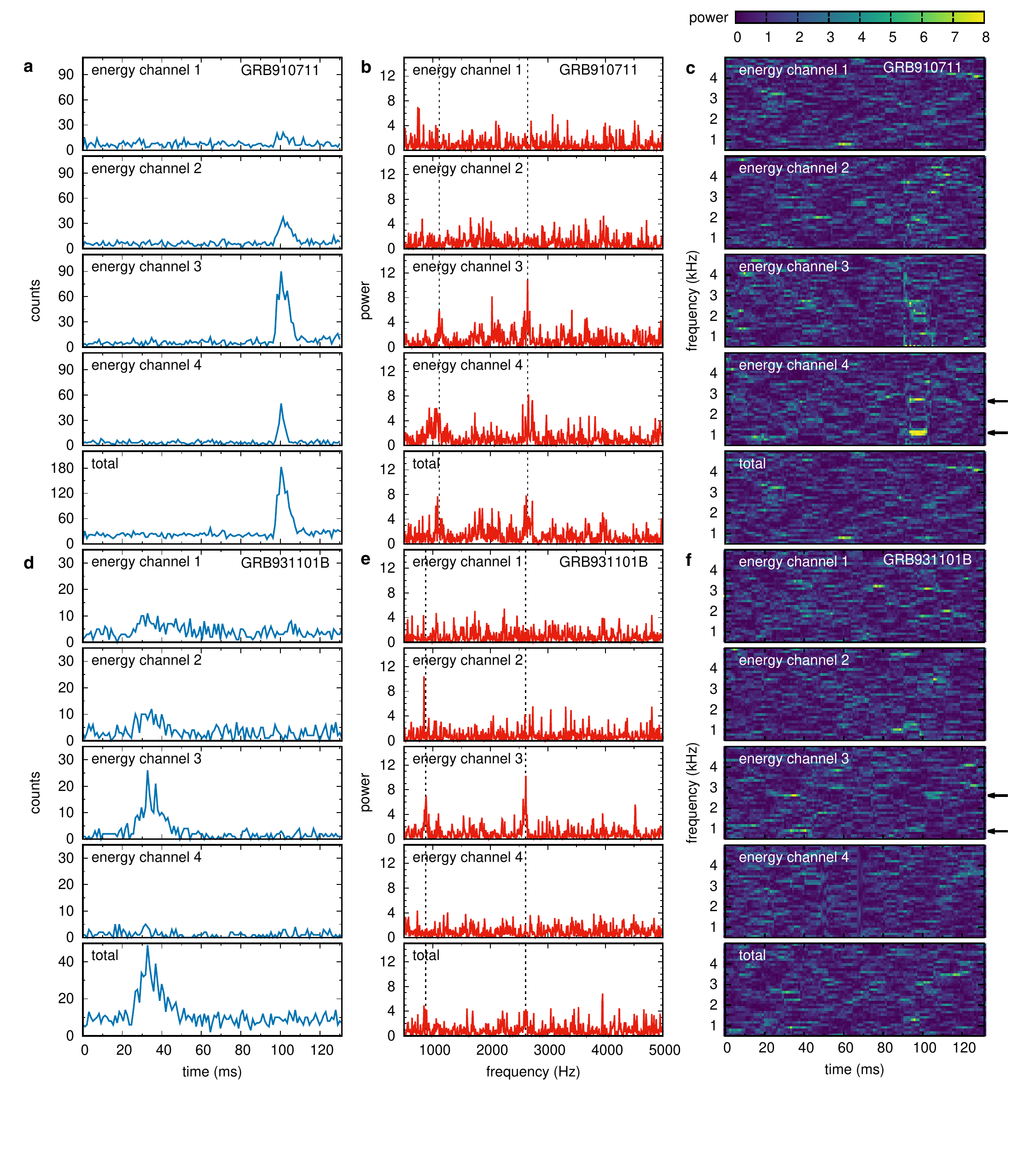}
\caption{$\vert$ \textbf{Energy dependence of burst properties.} \textbf{a,} Light curves of GRB~910711, in each of the four BATSE TTE channels and combined, over the segment of data (0.131072 s) that contains our strong signal.  We see that the higher-energy channels 3 and 4 have greater flux relative to the pre-burst background than the lower-energy channels 1 and 2.  \textbf{b,} The power spectra of GRB~910711, in each of the four BATSE TTE energy channels and combined.  The energy ranges are $20-50$~keV, $50-100$~keV, $100-300$~keV, and $>300$~keV for channels 1, 2, 3, and 4, respectively.  The vertical dotted lines show the centroid frequencies of the QPOs we identify in the summed channel 3 and 4 data (see Extended Data Table \ref{tab:bestfits}).  \textbf{c,} The spectrogram of GRB~910711, in each energy channel separately as well as in all channels combined, using the same intervals as in Figure~\ref{fig:spectrogram}.  The color bar on top of each set of plots shows the power scale.  The distribution of power, in time and frequency, is complicated. The black arrows indicate the mean values of the QPO frequencies given in Table \ref{tab:burstqpos}. In some cases (e.g., energy channel 3) there may be evidence for significant power prior to the main burst.  \textbf{d, e, f,} The same plots for GRB~931101B.}
\label{fig:energylightcurve}
\end{figure*}

In Extended Data Figure~\ref{fig:energylightcurve}, we show the light curves, power spectra, and spectrograms for each energy channel of each burst and for the sum of counts over all four BATSE TTE energy channels. As both GRBs are bright and hard, a small correction due to detector deadtime can be estimated \cite{2008GeoRL..35.6802G}, resulting in an effective deadtime of approximately $1\mu s$ per photon that is much smaller than the
period of our strongest QPO, $1/(2.6\, {\rm kHz}) \simeq 400\, \mu s$. Given that a particular burst is detected by $\sim 4$ of the eight BATSE detectors, this deadtime implies that we are missing approximately 2\% of the counts in GRB 910711 and 0.5\% of the counts in GRB 931101B on average during their whole duration (the percentage loss should be at most double these values at the peak of each burst). 

Although on balance the signals are strongest in the highest-energy channels 3 and 4, channel 2 in GRB~931101B shows a substantial signal near the lower-frequency QPO.  Moreover, examination of panels \textbf{c} and \textbf{f} of Figure~\ref{fig:energylightcurve} shows that the signals can appear over short times even before the main burst, and that the signals tend to be in the early phases of the bursts.  These properties may provide clues to the mechanism that makes the QPOs appear in gamma rays.

\bmhead{Data availability}
BATSE archival TTE data are available at \cite{BATSETTE}. 

\bmhead{Code availability}
Details about our codes have been published \cite{2019ApJ...871...95M}, but the code itself is not intended to be used publicly.

\backmatter

%\section*{References}

\section*{Declarations}

\bmhead{Acknowledgments}
We thank Brad Cenko, Alessandra Corsi, Liz Hays, Fred Lamb, Jay Norris, Luciano Rezzolla,  Nikhil Sarin, David Shoemaker and Zorawar Wadiasingh for discussions. 
 C. C. acknowledges support by NASA under award numbers 80GSFC17M0002 and TCAN-80NSSC18K1488.  M. C. M. was supported in part by NASA ADAP grant 80NSSC21K0649.
This work was partially conducted at the Aspen Center for Physics, which is supported by National Science Foundation grant PHY-1607611. 
Resources supporting this work were provided by the NASA High-End Computing (HEC) Program through the NASA Center for Climate Simulation (NCCS) at Goddard Space Flight Center.

\bmhead{Authors' Contributions}

C.C. led the project based on her idea of the possibility of a signal.  S.D. extracted the Fermi/GBM data and performed the galaxy host search.  A.L. extracted the Swift/BAT data and identified cosmic ray contamination in that data.  M.C.M. obtained the BATSE data and performed most of the data analysis.  R.P. provided expertise about possible systematic errors in the BATSE data.  All authors contributed ideas to the manuscript.

\bmhead{Competing interests}

The authors declare no competing interests (financial or non-financial).

\bmhead{Author information}

Correspondence and requests for materials should be addressed to  chirenti@umd.edu

\end{document}